\pdfoutput=1
\documentclass[conference,a4paper,10pt]{IEEEtran}
\usepackage[latin1]{inputenc}
\usepackage{times,amsmath}
\usepackage{amsfonts}
\usepackage{pstool}
\usepackage{subfigure}
\usepackage{multirow}
\usepackage{multicol}
\usepackage{enumerate}
\usepackage{graphicx}
\usepackage{mathtools, cuted}
\usepackage{MnSymbol}
\usepackage{stfloats}
\usepackage[table]{xcolor}
\usepackage[square, comma, sort&compress, numbers]{natbib}
\usepackage{nohyperref}
\usepackage{algorithm,algorithmic}
\usepackage{bigints}
\usepackage{amsmath}
\usepackage{adjustbox}
\usepackage{lipsum}

\usepackage{caption}

\usepackage{longtable}
\newcounter{tempEquationCounter}
\newcounter{thisEquationNumber}

\makeatletter
\newcommand{\vast}{\bBigg@{4}}
\newcommand{\Vast}{\bBigg@{5}}
\newcommand\numeq[1]%
  {\stackrel{\scriptscriptstyle(\mkern-1.5mu#1\mkern-1.5mu)}{=}}
\makeatother

\graphicspath{ {Figures/} }

\usepackage{acronym}

\begin{document}


\title{Underwater and Air-Water Wireless Communication: State-of-the-art, Channel Characteristics, Security, and Open Problems}
\author{
\IEEEauthorblockN{Waqas  Aman$^\ast$, Saif Al-Kuwari$^\ast$, Ambrish Kumar$^\dagger$,  Muhammad Mahboob Ur Rahman$^\ddagger$, and Muhammad Muzzammil$^\ast$ }
$^\ast$Division of Information and Computing Technology, College of Science and Engineering, \\Hamad Bin Khalifa University, Qatar Foundation, Doha, Qatar. \\$^\dagger$The Center for Telecommunication Research, School of Postgraduate Studies \& Research\\ Sri Lanka Technological Campus (SLTC), Sri Lanka \\
$^\ddagger$Electrical Engineering Department, Information Technology University, Lahore 54000, Pakistan. \\
$^\ast$waqasaman87@gmail.com,   $^\ast$smalkuwari@hbku.edu.qa,  $^\dagger$ambrish.nsit@gmail.com, \\
$^\ddagger$mahboob.rahman@itu.edu.pk,
$^\ast$mmuzzammil@hbku.edu.qa
}

\maketitle


\maketitle

\begin{abstract}
We present a first detailed survey on underwater and air-water (A-W) wireless communication networks (WCNs) that mainly focuses on the security challenges and the countermeasures proposed to date. For clarity of exposition, this survey paper is mainly divided into two parts. The first part of the paper focuses on the state-of-the-art underwater and A-W WCNs whereby we outline the benefits and drawbacks of the four promising underwater and A-W candidate technologies: radio frequency (RF), acoustic, optical and magnetic induction (MI), along with their channel characteristics. To this end, we also describe the indirect (relay-aided) and direct mechanisms for the A-W WCNs along with their channel characteristics. This sets the stage for the second part of the paper whereby we provide a thorough comparative discussion of a vast set of works that have reported the security breaches (as well as viable countermeasures) for many diverse configurations of the underwater and A-W WCNs. Specifically, we provide a detailed literature review of the various kinds of active and passive attacks which hamper the  confidentiality, integrity, authentication and availability of both underwater and A-W WCNs. Finally, we highlight some research gaps in the open literature and identify security related some open problems for the future work.
\end{abstract}




\section{Introduction}
\label{sec:intro}

Water covers 71$\%$ of the earth's surface (of which 96\% is held by oceans). Many activities, such as hunting for natural resources, trading, military operations, monitoring marine pollution, make the ocean a cardinal spot. Consequently, establishing communication between ground stations, airborne nodes (i.e., satellites, aeroplanes and unmanned aerial vehicles) and underwater sensor nodes, such as  submarine and  Autonomous Underwater Vehicles (AUVs),  are crucial to facilitate previously mentioned activities. Realizing its importance, over the last few decades, researchers have done a plethora of research in this area \cite{Kausal:Access:2016,Zeng:CST:2017,Gussen:JCIS:2016, Waqas:AINA:2016}.  

Over the years, radio frequency (RF), acoustic, and optical underwater wireless communication networks (WCNs) have gained widespread acceptance and adoption by the naval and oceanic communities. RF was used in early days of underwater wireless communication but lost its efficacy due to the severe attenuation it suffered from, which limited the effective range of communication. Long-range underwater wireless communication was also achieved using RF but with very large antennas \cite{WinNT}. Generally, acoustic underwater WCNs are preferred when long communication range is needed \cite{1090990}, but it is typically affected by multi-path propagation. In addition, the low speed of acoustic waves drastically reduces its rate and have prominent Doppler effect. At the same time, underwater wireless optical communication is becoming popular due to its high communication rate and relatively high speed. However, compared to acoustic, optical communication offers moderate communication range \cite{Kausal:Access:2016}. Moreover, magnetic induction (MI) based underwater WCNs provide Doppler free moderate rate (relatively higher than acoustic) and immune to multipath communication, but with moderate range due to conductivity of water medium \cite{li2019survey}. 

Recently, Air-Water (A-W) wireless communication, where some airborne nodes communicate with underwater nodes either directly or via relay/surface node, has attracted increasing attention from the research community. In indirect A-W communication, the surface node maintains two interfaces to independently communicate with the air and underwater nodes, then replays traffic between them. On the other hand, in direct A-W communication, the signals penetrate the water surface to make a direct link between air and underwater nodes. In this form of communication, RF and optics are commonly used on the over water part, with acoustic and optics are usually used for the underwater part \cite{Tonolini:SIGCOM:2018,Islam:ICC:2018}. MI communication has also been used for A-W direct communication due to same magnetic permeability in both air and underwater \cite{sun2010magnetic, domingo2012magnetic, li2019survey, zhang2019relay}. 


On the other hand, while wireless communication facilitates a much desired efficiency and portability, its broadcast nature inevitably makes it vulnerable to many types of attacks, such as eavesdropping, spoofing, denial of service, and jamming.  
In fact, attacks exploiting vulnerabilities in communication systems used for underwater and A-W wireless communication have already been demonstrated \cite{Kong:OE:17, Han:CommMag:2015,Domingo:WC:2011}. Today, securing these networks continues to be a demanding requirement that attracts the research community at large. 

In this survey, we provide a thorough and updated review of underwater wireless communication, from the underlying communication technologies and security aspects to the currently emerging open problems in this domain.  

\subsection{Contributions} The contributions of this survey can be summarized as follows:
\begin{itemize}
    \item This survey provides a thorough and updated review of the state-of-the-art of underwater wireless communications. 
    \item The previous survey papers discuss the underwater WCNs in one medium (i.e., water), and mostly consider acoustic communication technology. On the other hand, we review multiple technologies (i.e., acoustic, RF, optical, and MI) and scenarios (i.e., underwater WCNs, and A-W WCNs). 
    \item While some existing surveys \cite{lacovara:MTS:2008,Khalighi:ICTON:2014,Gussen:JCIS:2016,Kausal:Access:2016,Zeng:CST:2017,saeed:AHN:2019,Sajmath:ICSSS:2020,islam2022survey} mainly focus on the communication aspects of underwater WSNs (e.g., energy efficiency, routing, etc.), this survey further emphasizes on the security aspect of underwater and A-W WCNs, under diverse configurations and system models. Even though some surveys discussed security aspects of underwater WCNs, e.g., \cite{Domingo:WC:2011,Han:CommMag:2015,Lal:Ucomms:2016,YANG:phycom:2019,Jiang:COMMST:2018}, this paper is the first that focuses on the security aspect of both underwater and A-W WCNs. 
    \item In this survey, we identify a few gaps in the current literature and discuss in detail some open problems that we believe that the community should pay more attention to.
\end{itemize}
For clarity of exposition, Table \ref{tb:1} provides a crisp comparison between the earlier survey papers and our paper. 


 


\begin{table*}[htb!]
\centering

\begin{adjustbox}{{width=1\textwidth}}
\small
\begin{tabular}{ |c|c|c|c|c|c|c|}
 \hline

 \hline
 
\bf{Survey}  &  \multicolumn{4}{c|}{\bf{Security of Underwater Wireless Communication}} & \multicolumn{2}{c|}{\bf{Security of Air-Water Wireless Communication}}    \\

 &  RF & Acoustic & Optical & MI & Indirect &Direct  \\ \hline
\cite{Domingo:WC:2011} (2011) & $\times$ &$\checkmark$ &$\times$ &$\times$ &$\times$ &$\times$  \\
\hline
 \cite{Han:CommMag:2015} (2015)  &$\times$ &$\checkmark$ &$\times$ &$\times$ &$\times$&$\times$  \\
\hline
\cite{Lal:Ucomms:2016} (2016)  &$\times$ & $\checkmark$ &$\times$ &$\times$ & $\times$&$\times$ 
\\
\hline
\cite{Jiang:COMMST:2018} (2018)   &$\times$ &$\checkmark$ &$\times$&$\times$& $\times$ &$\times$  \\ \hline
\cite{YANG:phycom:2019} (2019)  &$\times$ &$\checkmark$ &$\times$&$\times$& $\times$ &$\times$
\\
 \hline
This Paper & $\checkmark$ &$\checkmark$ &$\checkmark$&$\checkmark$ &$\checkmark$ &$\checkmark$ 
\\
 \hline
\end{tabular}
\end{adjustbox}
\caption{Our survey paper is more comprehensive than the earlier survey papers as it covers various technologies and various scenarios to address security challenges faced by both underwater and A-W WCNs. }
\label{tb:1}
\end{table*}

\subsection{Organization} The rest of this paper is organized as follows. Section-\ref{sec:2} introduces underwater wireless communication with a brief discussion about the four widely used technologies: RF, acoustic, optical and MI communication. Section-\ref{sec:3} discusses air-water wireless communication and describes both  indirect and the newly developed domain of direct air-water wireless communication. Section-\ref{sec:4} reports on the security and state-of-the-art of underwater and air-water wireless communication, focusing on  authentication, confidentiality, and integrity. Finally, in Section-\ref{sec:5}, we conclude the paper and discuss open problems for future work.


\section{Underwater Wireless Communication}
\label{sec:2}
\begin{figure}[htb!]
\centering
\includegraphics[width=3.5in]{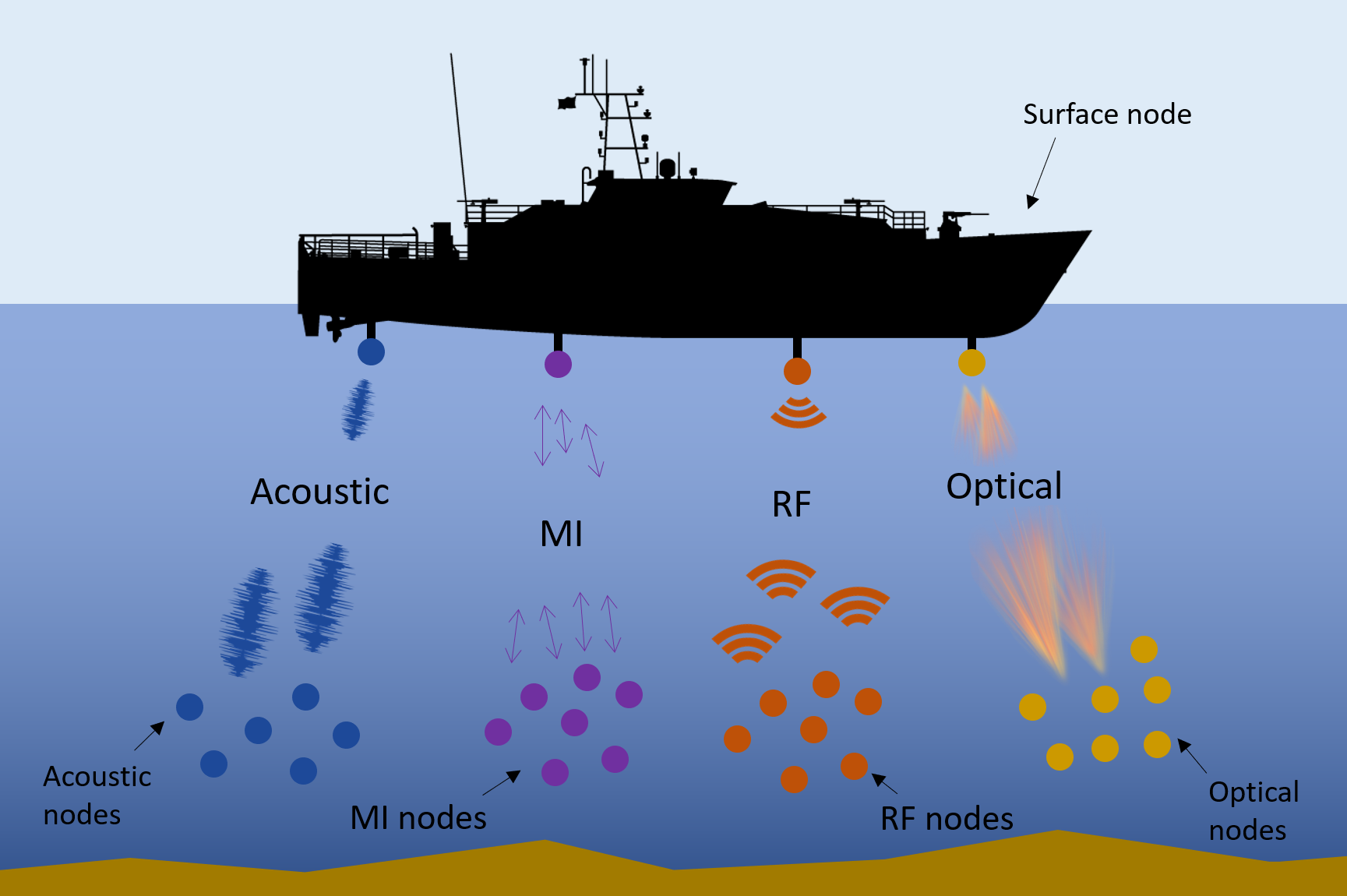}
\caption{Illustration of underwater WCNs.}
\centering
\label{fig:UWWCN}
\end{figure}

In underwater wireless communication, submarines, AUVs, and other sensor nodes are typically considered underwater nodes which are wirelessly connected with the surface nodes (e.g., ships, buoys) as illustrated in Fig. \ref{fig:UWWCN}. 

In the following subsections, we discuss the four most widely used technologies for underwater WCNs, namely: RF, acoustic, optical and magnetic induction (MI); Table \ref{tb:2} provides a brief comparison between these four technologies.  
\begin{table*}[ht!]
\centering
\begin{adjustbox}{width=1\textwidth}
\small
\begin{tabular}{ |c|p{4cm}|p{4cm}|p{2cm}|p{2cm}|p{2cm}|p{2cm}|}
 \hline

 \hline
\bf{System}  & \bf{Advantages} & \bf{Disadvantages}& \bf{Operating Frequency Range (Hz)}& \bf{Bandwidth} (Hz)&  \bf{Data rates}  & \bf{Deployment Example}  \\
 \hline
RF   & Fast carrier (velocity $\approx$ 225,000km/s), immune to acoustic noise,   & High attenuation, large transceivers, moderate latency, short range (less than 10m) & $30-300$ for ELF, MHz for others & $\approx$ 300 for ELF, MHz for others &1-2Mbps for 1-2m, 50-100bps for 200m &  ELF \cite{WinNT}  \\
  \hline
 Acoustic &  Travels over long distances (in kilo meters), small transceivers (compared to RF)  & Low data rates, low bandwidth, high noises, high latency, bulky transceivers (compared to optical) &$10-10^5$ & $\approx$ 10K&1.5-50Kbps for 0.5Km, 0.6-3Kbps for 28-120Km  &   Gertrude \cite{1090990}, Popoto acoustic modems \cite{Link1}, Aqua-scent modems \cite{Link2}  \\
 \hline
Optical & Fast carrier (velocity $\approx$ 225,000km/s, unlicensed spectrum, high bandwidth, secure, low latency, light transceivers  &  Needs line of sight due to high directivity of beam, moderate attenuation, medium range (100m)& $10^{12}-10^{14}$ & 10-150M  &1 Gbps for 2m, 1 Mbps for 25m  & LUMA by by Hydormea \cite{Link3}\\
  \hline
  MI & Fast carrier (velocity $\approx$ 225,000km/s, data rate in Mbps, low transmission delay, stable and predictable channel response, no multipath, no Doppler effect, stealth capabilities (non-audible and non-visible), and no impact on marine life & Medium range (10-100m), orientation sensitivity, and conductivity of seawater& $10^3-10^6$ & $\approx$ 2K  & Mbps & M$^2$I \cite{guo2017practical}, low-cost MI sensor nodes \cite{Ahmed2018design} \\
  \hline
\end{tabular}
\end{adjustbox}
\caption{Comparison of different underwater wireless communication technologies}
\label{tb:2}
\end{table*}

\subsection{RF-based Underwater WCN}
\label{subsec:RF}
RF-based underwater wireless communication networks offer many distinct advantages over their acoustic-based and optical-based counterparts; consequently, they have been used for various underwater applications~\cite{Moore@1967}. In fact, RF-based underwater WCNs have a rich history that goes back to the early years of $19^{\text{th}}$ century ~\cite{Siegel:TAP:1973}. One prominent example of a legacy RF-based underwater WCN is the so-called Extreme Low Frequency (ELF) communication system deployed by the US NAVY in Wisconsin state with an antenna's length exceeding 25 km \cite{WinNT}. This system enabled one-way communication with the submarines in order to (localize them, and thus) help them reach the surface of water safely. Indeed, due to its Point-to-Multipoint (P2MP) propagation nature, RF communication is suitable for cross-medium communication (air-water or water-to-air) where the terrestrial RF-based WCNs connect with the RF-based underwater WCNs ~\cite{Pompili@2009}. In contrast to their acoustic and optical counterparts, RF-based underwater WCNs are more tolerant to underwater turbulence and turbidity effects~\cite{Xianhui:ICM:2010}. This feature is certainly beneficial when underwater environmental conditions are random and/or unpredictable. Furthermore, RF signals can also propagate in adverse water conditions, while the optical waves are susceptible to particles and marine fouling~\cite{Palmeiro@2011}. Unlike acoustic communication, RF radiation is immune to acoustic noise, and it has no known effects on marine life~\cite{Gussen:JCIS:2016}.  

Despite the advantages RF-based underwater WCNs can offer, the heavy absorption loss at high frequencies and the requirement of excessively large antennas at low frequencies make this technology less attractive for deployment in defense and civilian marine underwater WCNs. Moreover, it also suffers from many other drawbacks, such as high attenuation (beside absorption and scattering, seawater conductivity seriously affects the propagation of electromagnetic waves), low bandwidth, very short communication range (typically less than 10m) and requiring large antennas. Nevertheless, occasionally, RF-based underwater WCNs have been practically deployed, as reported in the literature:  \cite{Xianhui:ICM:2010,jiang:JEAA:2011,qureshi:sensors:2016,Saini:IC3TSN,Soomro:RTSS:2017,Kenechi:APC:2019,Maher:ITCE:2019,Ganesh:WiSPNET:2019,ramdhan:BEEI:2019,Takizwa:JECNC:2021}. \\

\noindent\emph{RF Channel Characteristics}\\
\textit{Attenuation:}
The main factor of attenuation in the RF underwater channel is the absorption loss. The coefficient of absorption for seawater is given as:
\begin{align}
    \alpha_{\text{sw}}=\sqrt{\pi f \mu \sigma},
\end{align}
where $\sigma$ is the conductivity, $\mu$ is the permeability, and $f$ is the operating frequency. The absorption coefficient for fresh water is given as:
\begin{align}
    \alpha_{\text{fw}}=\frac{\sigma}{2}\sqrt{\frac{\mu}{\epsilon}},
\end{align}
where $\epsilon$ is the permittivity.\\
\textit{Transfer Function:} The transfer function of RF is given as \cite{zoksimovski2015underwater}:
\begin{align}
    H(f)=H_0e^{-\alpha(f)d}e^{-j\psi(f)},
\end{align}
where $H_0$ is a DC channel gain, $d$ is distance, $f$ is the operating frequency and $\psi(f)$ is the phase.

\subsection{Acoustic-based Underwater WCN}
Acoustic-based underwater wireless communication networks utilize acoustic modems (that consists of a sensitive acoustic receiver, i.e., hydrophone and an acoustic transmitter/projector) both at the surface node and at the underwater node ~\cite{Ian@2005},~\cite{Jiang@2018}. The current state-of-the-art in the design of acoustic modems with compact form-factor makes acoustic-based underwater WCNs appealing for many shallow water and deep water applications ~\cite{Zhang@2010}. Furthermore, the inherent ability of acoustic-based underwater WCNs to provide long-range transmission contributed to its widespread adoption for a plethora of marine applications, such as sound navigation and ranging (SONAR) protocols~\cite{SONAR:2003}. However, while the attenuation of acoustic waves at higher frequencies is still severe, the compact form-factor of the acoustic modems remains the preferred advantage, to the oceanic engineers. 

Although acoustic-based underwater WCNs have gained widespread adoption, they have certain intrinsic technical limitations as well~\cite{Dharrab@2013}. For example, the underwater acoustic frequencies vary between 10 Hz to 1 MHz, which results in low transmission rates. Furthermore, the acoustic link is significantly more affected by the water temperature, path-loss, noise, multi-path effect, and Doppler spread that might create a severe communication delay from the acoustic source to the acoustic receiver~\cite{Zhang@2010}. As a result, acoustic-based underwater WCNs cannot support the applications where real-time large volume data exchange is required. Moreover, their transceivers are bulky, costly, and energy-consuming and can also affect marine life~\cite{5425366}. \\

\noindent\emph{Acoustic Channel Characteristics} \\
\textit{Path-loss:}
The path-loss of line-of-sight (LoS) UWA channel between a node pair having distance $d$ separation is given as \cite{Stojanovic:1ACMIWUN:2006}

\begin{align}
\label{eq:pll}
\text{PL}(d,f) =  d^{\nu}   \alpha(f)^d. 
\end{align}

Alternatively, the dB scale representation of (\ref{eq:pll}) is

\begin{align}
\label{eq:pldb}
\text{PL}(d,f)_{\text{dB}} = \nu 10\log d + d \alpha(f)_{\text{dB}}. 
\end{align}

Basically, the  path-loss of LoS underwater acoustic (UWA) channel is the summation of spreading loss ($\nu 10\log d$) and absorption loss ($d \alpha(f)_{\text{dB}}$) in dB scale,
where $\nu$ is the spreading factor which takes $0$ or $1$ values, while  the  absorption coefficient $\alpha(f)_{\text{dB}}$ is given as \cite{Stojanovic:1ACMIWUN:2006}


\begin{align}
\alpha(f)_{\text{dB}}=\frac{0.11            
f^2}{1+ f^2}+\frac{44f^2}{4100+f^2}+2.75\times 10^{-4}f^2+0.003.
\end{align}



\begin{figure}[h!]
  \centering
  \subfigure[Absorption coefficient vs. frequency. \label{subfig:acoustic_absorptioncoeff}]{\includegraphics[scale=0.50]{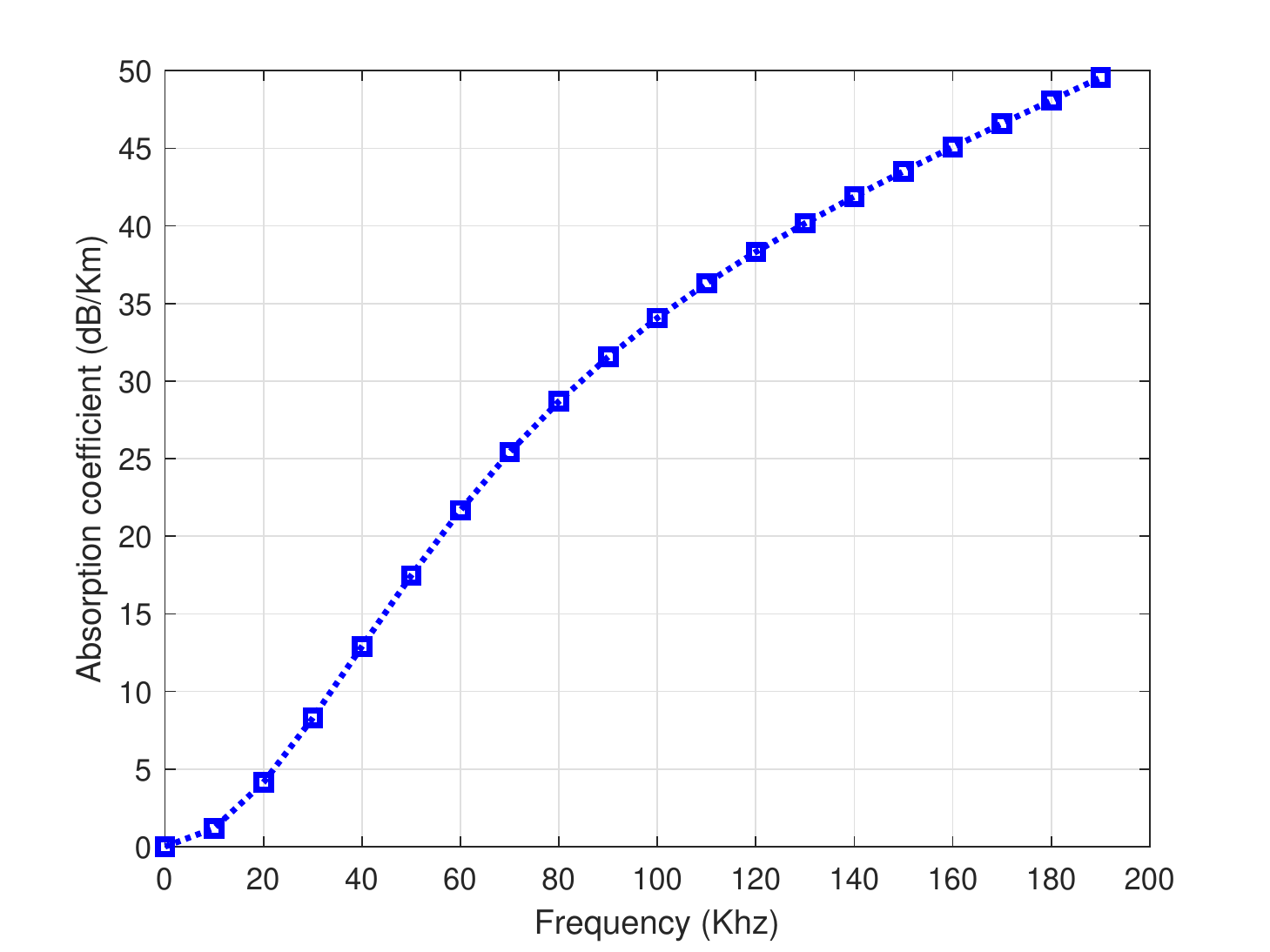}}
  \subfigure[Path-loss vs. transmission range under different frequencies. \label{subfig:acoustic_pl_frequency}]{\includegraphics[scale=0.50]{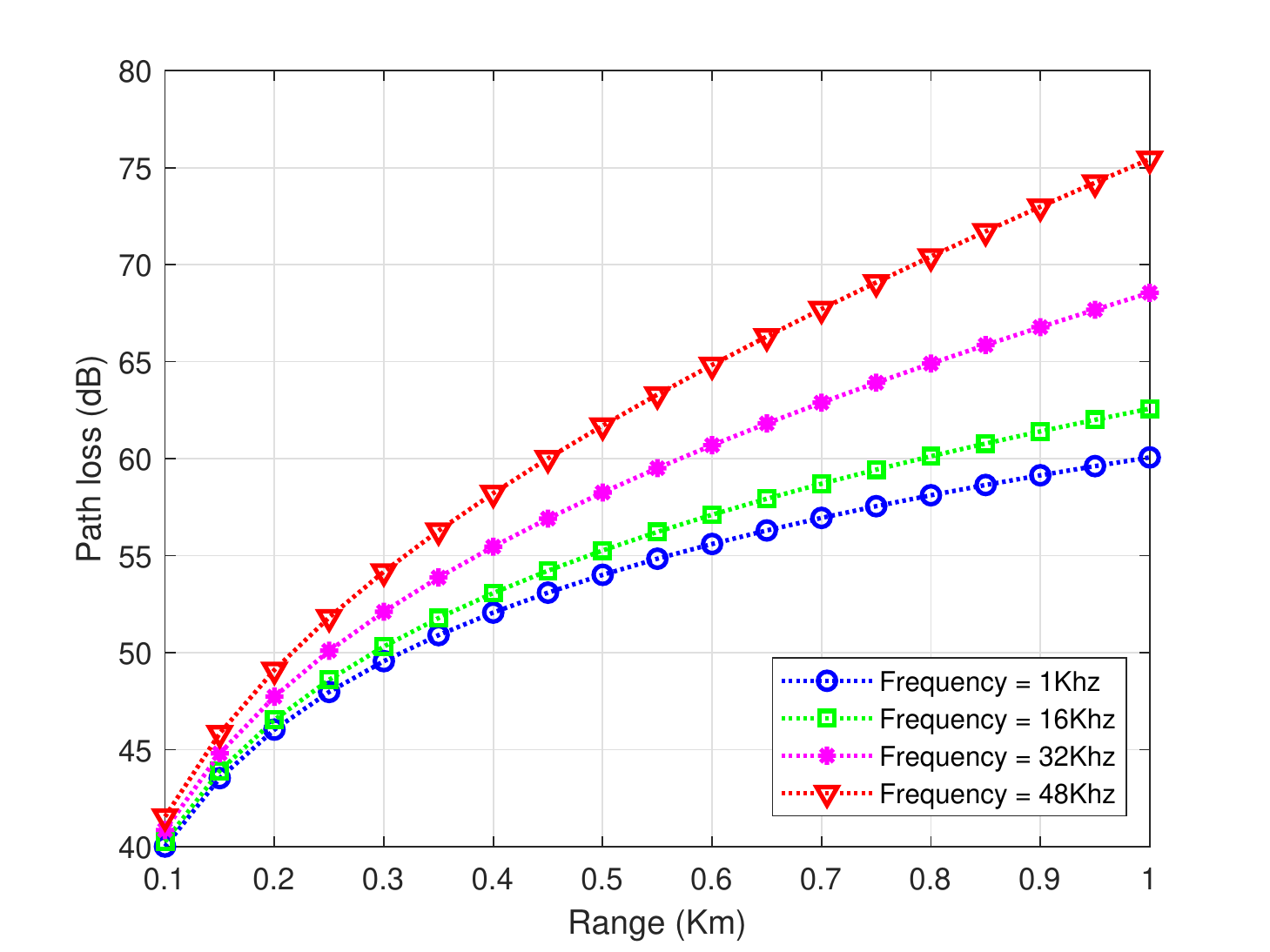}} 
 \caption{Acoustic communication: path-loss vs. transmission range.} 
\label{fig:pl_acoustic}
\end{figure}
From Fig. \ref{fig:pl_acoustic}, simulations of absorption coefficient versus frequency and path-loss versus transmission range are carried out. It can be depicted from Fig. \ref{subfig:acoustic_absorptioncoeff} that absorption coefficient increases by increasing frequency.  While, from Fig. \ref{subfig:acoustic_pl_frequency}, increasing trend of path-loss can be seen by increasing frequency and transmission range at the same time. At a transmission range of 1 Km, the path-loss increases by 15 dB when frequency changes from 1 KHz to 48 Khz.

 \textit{Noise:}
There are mainly four sources of noise in the UWA channel: thermal noise, wave noise, shipping noise, and turbulence noise. We discuss these in more detail  below:
\begin{itemize}
    \item 
{Thermal noise:}
Thermal noise is due to thermal agitation of molecules and operates above the $100$ kHz {(Note that the noise effect is zero outside the given  range of frequencies)}, and its power spectral density (PSD) in dB is given as \cite{Milica:PSD:2007}

\begin{equation}
    N_{\text{T}}(f)=-15-20\log(f).
\end{equation}
\item {Wave noise:}
Wave noise is the most prominent component among the four mentioned noises because it exists in the desired frequency range (i.e., $100$ Hz to $100$ kHz). This produces due to the surface movement of the sea, which is induced by the wind. The PSD of wave noise in dB is given as

\begin{equation}
    N_\text{W}(f)=50+20\log(f)-40\log(f+0.4)+7.5\sqrt{w},
\end{equation}

where $w$ is the speed of the wind.
\item{Shipping noise:}
The noise generated due to shipping activities is shipping noise. This noise exists in the range $10<f<100$ Hz. The PSD of shipping noise is given as

\begin{equation}
    N_\text{S}(f)=40-60\log(f+0.3)+26\log(f)+20(s-0.5),
\end{equation}

where $s$ represents the shipping activities (i.e. $s=0(1)$ for low(high)).
\item {Turbulence noise:}
Turbulence noise is generated due to random movement of sea surface due to wave currents, and it exists at very low frequencies, i.e., $f<10$ Hz. The PSD  of Turbulence noise is given as

\begin{equation}
    N_{\text{Tr}}(f)=17-30\log(f).
\end{equation}
\item {Net noise:}
The PSD of net noise can be expressed as

\begin{equation}
N_{\text{NN}}(f)=N_{\text{T}}(f)+N_{\text{W}}(f)+N_{\text{S}}(f)+N_{\text{Tr}}(f).
\end{equation}
\end{itemize}
There is a very good approximate of PSD of net noise in the range $1<f<100$ KHz and is given as \cite{Stojanovic:1ACMIWUN:2006}  

\begin{align}
\label{eq:noise}
N(f)_{\text{dB}} \approx N_1- \tau 10\log (f),  
\end{align} 

where $N_1$ and  $\tau$ are the experimental constants.\\
\textit{Multi-path:}
Now, let $H_j$ be the channel gain of UWA channel for $j^{th}$ sub-carrier. Then during $k^{\text{th}}$ block/time-slot, $H_j(k)$ which comprises total $L$ paths can be expressed as

\begin{equation}
\label{CGinMP}
H_j(k)=\sum_{l=1}^L \frac{1}{\sqrt{\text{PL}(d_\text{l},f_j)}} h_\text{l}(k) e^{-j2\pi f_j \xi_l(k)}, 
\end{equation}

where $\text{PL}(d_\text{l},f_j)$ is the path-loss of the $l^{\text{th}}$ path having distance $d_\text{l}$, and the delay of $l^{\text{th}}$ path is $\xi_\text{l}(k)=\frac{d_l}{v}$, where $v$ is the underwater speed of acoustic wave. $h_\text{l}(k)$ is the $l^{\text{th}}$ path gain which is modelled as independent, first-order auto-regressive process.


\subsection{Optical-based Underwater WCN}
Optical-based underwater wireless communication networks utilize (visible or invisible) light to carry information whereby a laser diode (LD) and/or a light emitting diode (LED) are used as the sources for transmission~\cite{1093200,Alouini:2018,Huang:COL:19,Yan:IPJ:2019}. The optical underwater WCNs recently received attention due to their favorable characteristics such as high speed, high data rates, and high bandwidth \cite{Alouini:2018,Huang:COL:19,Yan:IPJ:2019,SPagnolo:Sensors:2020,Zhu:PQE:2020,Sun:JLT:20,Sait:IPJ:2021,Khan:IBCAST:2021,Lu:ITC:2021}. The effective communication range of optical-based underwater WCN is greater than its RF-based counterpart but quite lower than its acoustic-based counterpart (typically, up to 100m). It can potentially solve the problem of broadband and low-latency submarine WCNs~\cite{Zeng:CST:2017,SPagnolo:Sensors:2020,Zhu:PQE:2020}. The optical-based underwater WCNs can provide a high-speed data transmission rate (in Gbps) for a moderate transmission range~\cite{Kausal:Access:2016,Sun:JLT:20,Sait:IPJ:2021}. Moreover, their high-speed transmission also guarantees that optical-based underwater WCNs can be used for many real-time underwater applications, e.g., underwater video transmissions~\cite{7031910,Khan:IBCAST:2021,Lu:ITC:2021}. Even though optical-based underwater WCNs offer the great advantage of high data rates for short and medium ranges, its LoS requirement might sometimes be difficult to satisfy. In other words, misalignment between the orientation of the optical transmitter and optical receiver might deteriorate the performance of an optical underwater WCN.  \\

\noindent\emph{Optical Channel Characteristics} \\
\textit{Path-loss:}
The path-loss of optical channel as a function of wavelength $\lambda$ and distance $d$ can be expressed as
\begin{align}
    \text{PL}_{O}(\lambda,d)=\exp{(-e(\lambda) d)},
\end{align}
where $e(\lambda)$ is known as optical beam extinction coefficient which comprises of scattering coefficient $c(\lambda)$ and absorption coefficient $a(\lambda)$ expressed below;
\begin{align}
   c(\lambda) = a(\lambda)+s(\lambda).
\end{align}

The absorption coefficient can be computed as \cite{Kausal:Access:2016}:
\begin{align}
\label{eq:optical_att_coeff}
    a(\lambda)=\begin{array}{c} lim\\ \Delta r\rightarrow 0 \end{array}\frac {\Delta A\left ({\lambda }\right )}{\Delta r}=\frac{d (A(\lambda))}{dr},
\end{align}
where $A(\lambda)=\frac{P_\text{a}(\lambda)}{P_\text{i}(\lambda)}$ (known as absorbance),  ${P_a(\lambda)}$ is the absorbed power while ${P_\text{i}(\lambda)}$ is the incident power. Similarly, the scattering coefficient can computed as \cite{Kausal:Access:2016}:
\begin{align}
\label{eq:optical_sca_coeff}
    s(\lambda)=\begin{array}{c} lim\\ \Delta r\rightarrow 0 \end{array}\frac {\Delta S\left ({\lambda }\right )}{\Delta r}=\frac{d (S(\lambda))}{dr},
\end{align}
where $S(\lambda)=\frac{P_\text{s}(\lambda)}{P_\text{i}(\lambda)}$ (known as scatterance), and $P_\text{s}(\lambda)$ is the scattered power. In Eq. \ref{eq:optical_att_coeff} and \ref{eq:optical_sca_coeff}, the absorption and scattering coefficient are calculated by opting limit as thickness and $\Delta r$ is infinitesimally small \cite{Kausal:Access:2016}.\\
\begin{figure}[htb!]
\centering
\includegraphics[scale=0.5]{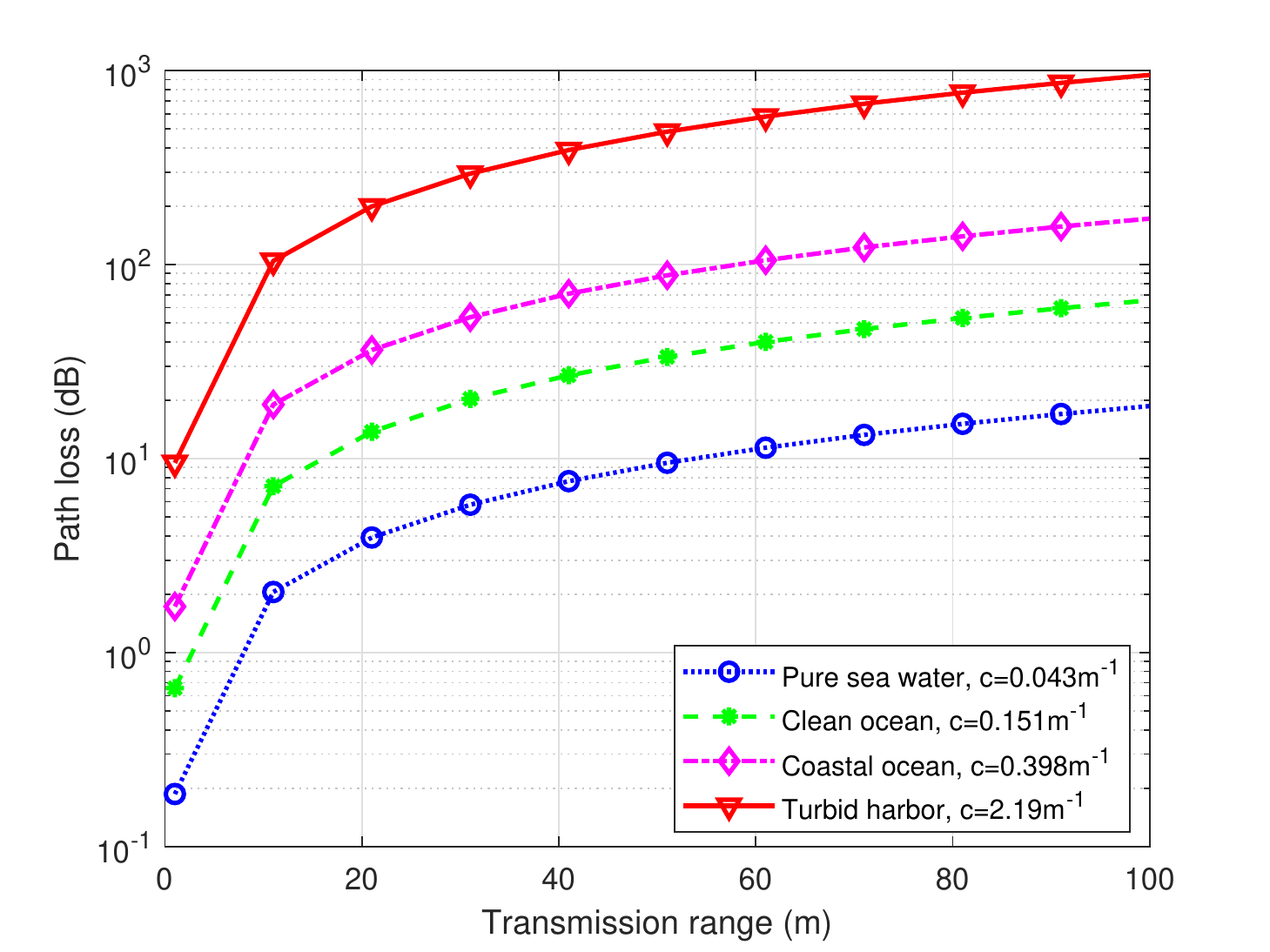}
\caption{Optical Communication: path-loss vs. transmission range.}
\centering
\label{fig:pl_optical}
\end{figure}
\textit{Noise:} The total noise can be expressed as:
\begin{align}
    P_{\text{noise}}=P_{\text{solar}}+P_{\text{blackbody}},
\end{align}
where $P_{\text{solar}}$ is noise power due to solar while $P_{\text{blackbody}}$ is noise power due to blackbody radiation.
\begin{align}
    P_{\text{solar}}=A_\text{R}\Delta\lambda T_\text{F} R_{\text{solar}}(\pi \text{FOV})^2 
\end{align}
where $\text{FOV}$ is field of view, $A_\text{R}$ is the receiver area, $\Delta\lambda$ is the optical filter width, $T_\text{F}$ is the transmissivity of optical filter and solar radiance $R_{\text{solar}}$ is given as:
\begin{align}
\label{eq:R_sol}
    R_{\text{solar}}=\frac{i_\text{r}R L_\text{f} \exp{(-K\delta})}{\pi},
\end{align}
where $i_\text{r}$ is the down-welling irradiance, $R$ is the reflectance of down-welling irradiance, $\delta$ is the underwater depth, $K$ is the coefficient of attenuation and $L_\text{f}$ is the
factor describing the directional dependence of underwater radiance. The noise form blackbody radiation can be expressed as:
\begin{align}
    P_{\text{blackbody}}=\frac{2hc^2 \gamma A_\text{R}\Delta\lambda T_\text{F} T_\text{A}(\pi \text{FOV})^2 }{\lambda^5(\exp{(\frac{hc}{ \lambda kT})}-1)},
\end{align}
where $h$ is a Planck constant, $c$ is the speed of light in water, $\gamma$ is the absorption factor,  $T_\text{A}$ is the transmission and $k$ is the Boltzmann constant.
\\ 
\textit{Channel Impulse Response:} According to \cite{6685978} , the CIR of underwater optical communication can be expressed in closed form as 
\begin{align}
    h(t)=C_1\Delta t \exp{(-C_2 \Delta t)}+C_3\Delta t \exp{(-C_4 \Delta t)},  \ \ \ (t>t_0)
\end{align}
 where $\Delta t=t-t_0$, $t$ is time scale while $t_0=\frac{d}{v}$ is the ratio of distance and velocity of light in water (also known as propagation time), $C_1,C_2,C_3,C_4$ are the four variables/parameters which can be found through least square approach. 
 \subsection{MI-based Underwater WCN}
 \label{subsec:MI_UWCNs}

In magnetic induction (MI)-based underwater wireless communication networks, the communication between any two nodes is performed by applying time varying signal to the transmitter (Tx) coil which generates magnetic field around its origin. When receiver (Rx) coil comes in the vicinity of that generated magnetic field, both Tx and Rx coils coupled with each other and information is exchanged \cite{muzzammil2020fundamentals}. MI-based underwater wireless communication is believed to be an alternative to RF, acoustic, and optical technologies in medium range underwater applications due to many advantages such as low transmission delay, moderate data rates (in Mbps), no multipath and Doppler effect, and stealth operation capabilities (non-audible and non-visible) \cite{muzzammil2020fundamentals, li2019survey}. Further, smooth air-to-water and water-to-air communication can be achieved with MI-based communication due to same magnetic permeability in both air and water \cite{sun2010magnetic, domingo2012magnetic}. The MI transceivers cost is much lower as compared to RF transceivers and acoustic modems and MI technology has the capability of underwater wireless power transfer as well \cite{muzzammil2020fundamentals}. Moreover, in contrast to acoustic and optical technologies, MI-based underwater wireless communication has no impact on the marine life. Besides these wonderful advantages of MI technology, orientation sensitivity (misalignment between Tx \& Rx coils) and conductivity in seawater can significantly affect the MI communication performance \cite{li2019survey, muzzammil2019directivity}. The magnetic field generated by Tx coil is directional in nature, and when Tx and Rx coils are not perfectly aligned to each other, the performance of MI-based WCNs will be significantly reduced. Further, to realize the practical implementation of MI-based WCN, the range of MI communication needs to be improved. The range is dependent on the magnetic field strength (MFS) and solutions such as increasing magnetic moment \cite{muzzammil2020fundamentals}, coil designing \cite{guo2017practical, sharma2017analytical, Ahmed2018design}, and relaying (waveguide, active, and hybrid) \cite{sun2010magnetic, domingo2012magnetic, li2019survey, qiao2020experimental} are introduced to maximize the MFS. \\

\noindent\emph{MI Channel Characteristics}

The MI channel modeling is mainly performed by two methods: equivalent circuit analysis \cite{sun2010magnetic, domingo2012magnetic} and magnetic field analysis \cite{gulbahar2012communication, guohongzhi2015channel}. The widely adopted method in the existing literature for calculating MI channel path-loss is equivalent circuit analysis in which MI transceivers are modeled either as a transformer model \cite{sun2010magnetic} or a two-port network model \cite{domingo2012magnetic}.

The total path-loss in MI based underwater WCNs comprises of two main components, which are path-loss due to MI and path-loss due to eddy currents. The total path-loss can be expressed as
\begin{align}
\label{eq:pl_MI_total}
    \text{PL}_{\text{MI}_{\text{T}}}=\text{PL}_{\text{MI}}+\text{PL}_{\text{EC}},
\end{align}

where the path-loss due to MI in dB scale can be expressed as \cite{domingo2012magnetic}
\begin{align}
\label{eq:pl_MI}
  \text{PL}_{\text{MI}}=-10\log \frac{{{R}_{\text{L}}}{{\omega }^{2}}{{M}^{2}}}{{{R}_{\text{Tx}}}{{({{R}_{\text{L}}}+{{R}_{\text{Rx}}})}^{2}}+{{R}_{\text{Tx}}}{{({{X}_{\text{L}}}+\omega {{L}_{\text{Rx}}})}^{2}}},  
\end{align}
where $\omega =2\pi f$ denotes the angular frequency, $R_{\text{Tx}}$ and $R_{\text{Rx}}$ are the Tx and Rx resistances, $R_{\text{L}}$ and $X_{\text{L}}$ are the load resistance and inductive reactance, $L_{\text{Rx}}$ denotes Rx inductance and mutual induction $M$ between transmitter and receiver can be represented as \cite{finkenzeller2010rfid}
\begin{align}
  M={\mu \cdot\pi \cdot N_{\text{Tx}}\cdot r_{\text{Tx}}^{2}\cdot N_{\text{Rx}}\cdot r_{\text{Rx}}^{2}\over 2\sqrt{\left(r_{\text{Tx}}^{2}+d^{2}\right)^{3}}}, 
\end{align}
where $\mu=\mu_0 \cdot \mu_r$ is the magnetic permeability, here $\mu_0=4\pi \cdot 10^{-7}H/m$ is the magnetic constant and $\mu_r=1$ is the relative permeability of water. Further, $N_{\text{Tx}}$ and $N_{\text{Rx}}$ are the number of turns in the Tx and Rx coils, $r_{\text{Tx}}^{2}$ and $r_{\text{Rx}}^{2}$ represent the Tx and Rx coils radii and $d$ is the distance between Tx and Rx coils.
 The path-loss due to eddy current can be written as 
 \begin{align}
 \label{eq;pl_EC}
    \text{PL}_{\text{EC}}=20\log ({{e}^{\alpha d}})=8.69\alpha d,
 \end{align}
 where $\alpha =1/\delta =\sqrt{\pi f\mu \sigma }$  denotes the attenuation which is inverse of the skin depth $\delta$, here $\sigma$ is the conductivity of the sea water.
 
 \begin{figure}[h!]
  \centering
  \subfigure[path-loss vs. transmission range under different frequencies when coil radius = 1.0m, number of turns = 500, conductivity of fresh water (FW) = 0.01S/m, and conductivity of sea water (SW) = 4.0S/m. \label{subfig:pl_frequency}]{\includegraphics[scale=0.50]{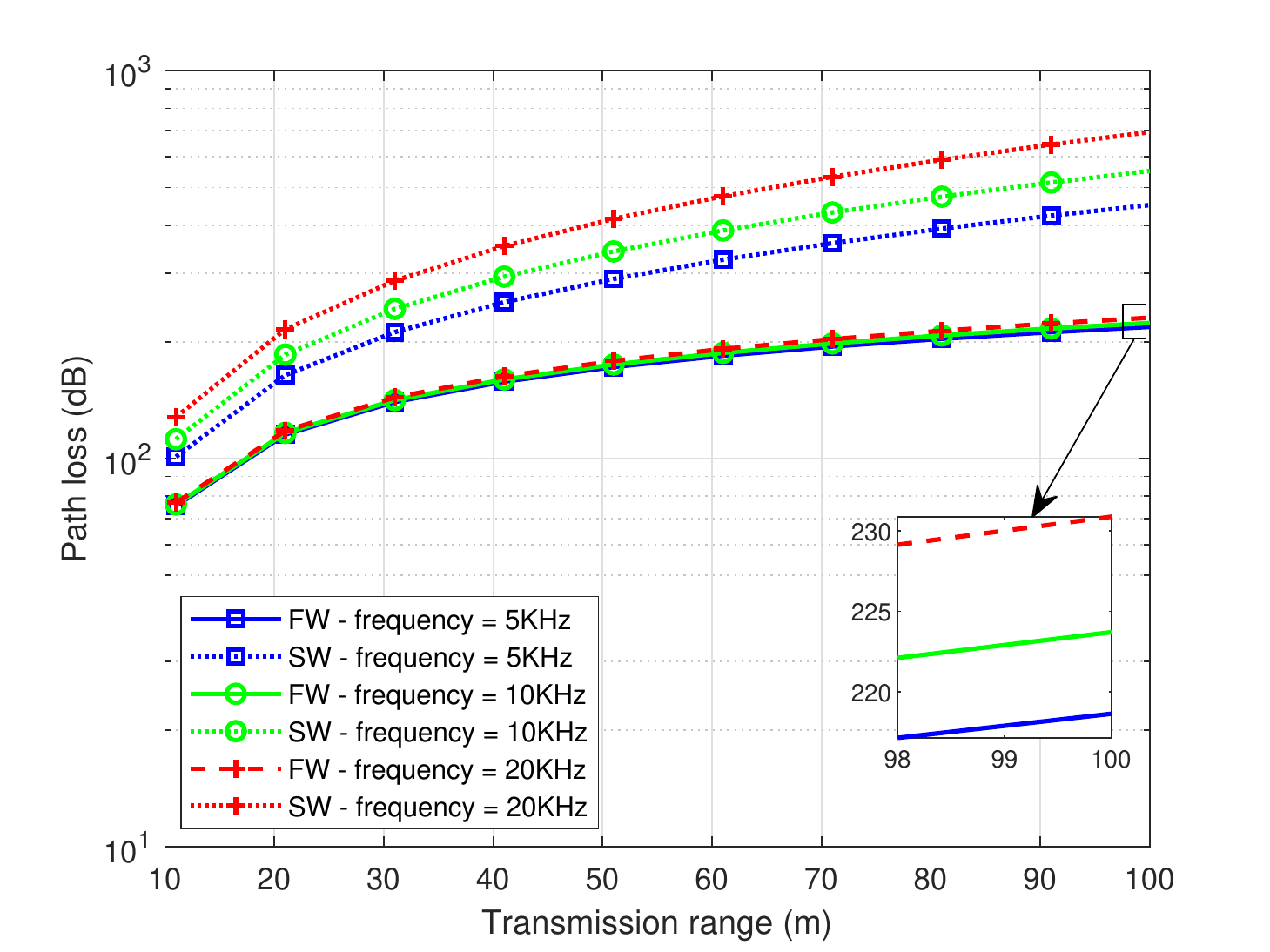}}
  \subfigure[path-loss vs. transmission range under different coil radius when frequency = 5KHz, number of turns = 500, conductivity of fresh water (FW) = 0.01S/m, and conductivity of sea water (SW) = 4.0S/m. \label{subfig:pl_coilradius}]{\includegraphics[scale=0.50]{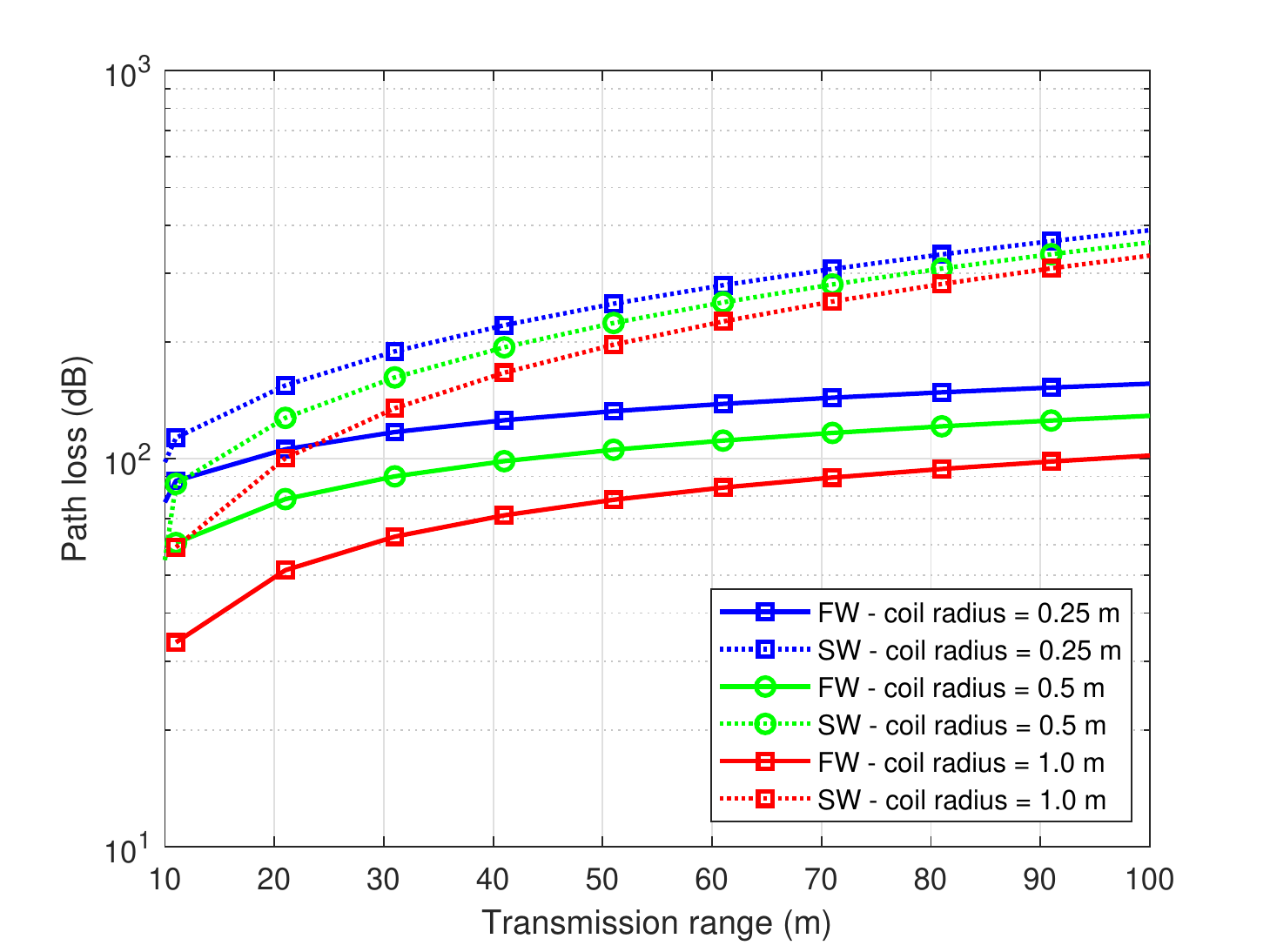}} 
 \caption{MI communication: path-loss vs. transmission range.} 
\label{fig:pl_MI}
\end{figure}
From Fig. \ref{fig:pl_MI}, simulation results of path-loss vs. transmission range are shown. Clearly, an increasing trend of path-loss can be seen against increasing frequency and transmission range as shown in Fig. \ref{subfig:pl_frequency}. Further, path-loss in fresh water is minimum as compared to the sea water due to conductive nature of the sea water. Fig. \ref{subfig:pl_coilradius} shows path-loss vs. transmission range under different coil radius. It can be depicted from Fig. \ref{subfig:pl_coilradius} that path-loss can be reduced by increasing the coil radius. Further, a similar trend of increased path-loss in sea water can be seen as compared to fresh water under same coil radius and transmission range.

It is believed that the noise in MI underwater WCNs is only thermal noise due to random motion of electron in resistive elements in the receiver circuitry, and the power of  thermal noise can be be expressed as \cite{6554972}:
\begin{align}
    N_\text{MI}\approx B K T,
\end{align}
where $B$ is the bandwidth, $K$ is the Boltzmann's constant and $T$ is the temperature in Kelvin. Further, the existence of ferro-magnetic materials may also leads to adding an additional noise, however, it strongly depends on the placement of these materials between the MI transceivers. Because, the placement of these materials either results in reduction or further improvement the magnetic field strength at the receiver end \cite{ahmed2016effects}.

\section{Air-water wireless communication}
\label{sec:3}
In this section, we provide a brief overview of A-W WCNs. A total of five configurations are considered, namely: 1) acoustic (in water) to RF (in air) \cite{Tonolini:SIGCOM:2018}, 2) acoustic (in water) to optical (in air) \cite{Zhong:ICM:2021}, 3) light (in water) to RF (in air) \cite{Anees:VTC:2019}, 4) direct light in both mediums (water and air) \cite{Carver:MCC:2021}, and 5) magnetic field in both mediums (water and air) \cite{zhang2019relay, tian2020test}. Typically, air-water wireless communication can either be indirect or direct depending on the existence of a relay/surface node.

\begin{figure*}[htb!]
\centering
\includegraphics[width=7in]{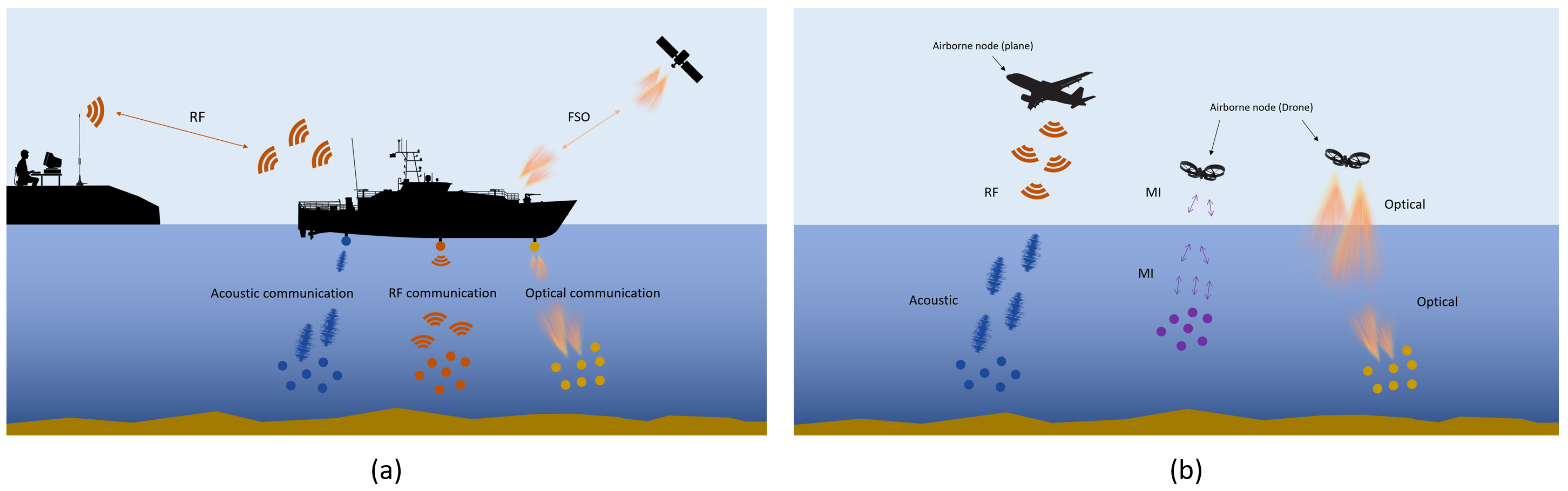}
\caption{Illustration of A-W WCNs. (a) Indirect (Surface node-assisted) communication, (b) Direct communication.}
\centering
\label{fig:SM}
\end{figure*}

\subsection{Indirect Air-Water Wireless Communication}
In indirect air-water wireless communication systems (also called surface node assisted communication), underwater nodes communicate with surface nodes (also called sink nodes), that are further connected to ground stations via satellite, microwave, or Free-Space Optical (FSO) links. The sink node is equipped with two transceiver interfaces: one for submerged nodes and the other for the water surface nodes. An illustration of such systems is provided in Fig. \ref{fig:SM}(a), where different underwater systems communicate with the surface node via a sink node attached to a ship. Consequently, the addition of an extra hop always leads to increased end-to-end delay, which, in turn, exposes the system to more attacks. Typically, optics in water, and RF in the air is considered for end-to-end analysis in indirect A-W communication \cite{Anees:VTC:2019}. However, recently, an experimental study on photoacoustic communication has been conducted in \cite{Zhong:ICM:2021}, where acoustic is used for underwater and optical communication is used for over water. MI technology is also recently utilized for A-W wireless communication that can work both ways: indirect and direct. In the case of MI-based indirect A-W wireless communication, there can be two different configurations. In configuration 1, the airborne, sink and underwater nodes are equipped with MI transceivers. The transmission/reception of information can be performed through the sink node which will act as a waveguide (relay) node between the airborne and underwater nodes. This configuration can be helpful in delay sensitive and long transmission range (as compared to the conventional MI technology) applications. While in configuration 2, the underwater node communicates with the sink node through MI transceivers and sink node further communicates with airborne nodes either through RF or FSO as depicted in Fig. \ref{fig:SM}(b). This configuration indeed can also be useful in delay sensitive and long range applications however at the cost of adding additional transceiver modules to the sink nodes which leads to an increased cost, weight and size of the sink node.\\

\noindent\emph{Indirect A-W Channel Characteristics} \\
The Channel Characteristics for the underwater part (i.e., RF, acoustic, optical, and MI) is discussed in Section \ref{sec:2} while for the aerial part (i.e., RF and Optical) terrestrial model for path-loss, noise and fading are applicable. The main aim of such systems is to analyze the end to end communication. However, in the case of configuration 1 of indirect MI A-W wireless communication, the channel characteristic model will be different as discussed in subsection \ref{subsec:MI_UWCNs} and it is presented below. \\
\textit{Channel Characteristics of MI-based Indirect A-W wireless communication-Configuration 1}: In this configuration, the sink node is used as a waveguide (relay node), hence, a conventional MI channel path-loss model is not applicable. Instead, MI-based waveguide channel model needs to be adopted, in which path-loss expression can be calculated by equivalent circuit analysis method based multi-stage transformer model \cite{sun2010magnetic} or multi-port network model \cite{domingo2012magnetic}. \\
\textit{Path-loss:} The path loss model for configuration 1 can be represented as \cite{domingo2012magnetic}

    \begin{align}
      \text{PL}_\text{MI-WG}=-10\log \bigg(\frac{{{R}_\text{L}}{{\omega }^{2}}{{M}^{2}}}{R\cdot {{({{R}_\text{L}}+R)}^{2}}}\cdot \text{ }\prod\limits_{g=a\;k=b}^{g=n-1\;k=n}{\frac{{{\omega }^{2}}{{M}^{2}}}{{{\left( 2R+\frac{{{\omega }^{2}}{{M}^{2}}}{|{{\gamma }_{g+1\;k+1}}|} \right)}^{2}}}}\bigg), 
    \end{align}
where ${P}_\text{Tx}$ and ${P}_\text{Rx}$ are the transmit and receive powers, $\omega$ is the angular frequency, and $M$ is the mutual coupling between two adjacent nodes (airborne, sink or underwater). The value of $\gamma_{g+1\;k+1}$ can be computed from appendix B of \cite{domingo2012magnetic}.

\subsection{Direct Air-Water Wireless Communication}
\label{SC:DA-W}
Surface nodes are usually attached to large buoys or ships, which can be easily spotted by the enemies/adversaries through the use of radars or telescopes. Therefore, in order to avoid the need for a surface floating node, a new (potentially revolutionary) paradigm has recently been proposed in~\cite{Tonolini:SIGCOM:2018}. In this first work, direct communication between underwater and over water nodes takes place, where acoustic communication was used for the underwater part and RF communication for over water part. The basic idea of direct air-water communication is illustrated in Fig. \ref{fig:SM}(b), where the airborne node receives and decodes information from the backscattered RF waves after striking the ripples produced by the acoustic waves on the surface. Similarly, the use of direct light from air to water has been explored in \cite{Islam:ICC:2018} and \cite{Islam:Access:2019}, followed by various works attempting to enhance the performance \cite{Nabavi:ICC:2019,Chen:OFC:2021,Lin:IPJ:2021,Shao:OFC:2021,Sun:OE:2019,Sun:Access:2020,Enhos:SECON:2021, (Watson:IJECE:2021, Guo:UCMMT:2019, Carver:MCC:2021,Moniara:RST:2020,Luo:ICCC:2019}. Moreover, as previously mentioned, MI technology is capable of direct A-W wireless communication due to the same magnetic permeability in air and water. In this case, both airborne and underwater nodes need to be equipped with MI transceiver as shown in Fig. \ref{fig:SM}(b). \\

Three different systems are considered for direct A-W communication, namely: Acoustic-RF based, Optical based and MI based. We now discuss the channel characteristics of all the three systems below.  
\\

\subsubsection{System 1: Translation Acoustic-RF (TARF) Communication} 
 The end to end channel of TARF comprises of below three parts. \\
 
\paragraph{Air part} The attenuation in this part is given as:
\begin{align}
\text{PL}_{\text{A}}^{\text{TARF}}=\frac{1}{2d_0}
\end{align}
where $d_0$ is the distance from airborne node to water surface.

\paragraph{Surface part} The attenuation at the A-W interface is given as:
\begin{align}
    \text{PL}_{\text{A-W}}^{\text{TARF}}(\omega, t)= \frac{P(\omega, t)}{\rho \omega v}, 
\end{align}
where $\omega$ is the angular frequency, $t$ is time, $\rho$ is the water density and $P$ is the overall acoustic pressure. Besides, the sensed power at this interface for a given incident power $P_{\text{i}}$ is given as
\begin{align}
P^{\text{TARF}}_{(\text{A-W})_{\text{sensed}}}= \frac{P_{\text{i}}}{\rho v \omega^2},
\end{align}

\paragraph{Underwater part} The path-loss/attenuation in water medium for TARF is given as
\begin{align}
    \text{PL}_{\text{UW}}^{\text{TARF}}= \frac{\exp{(\gamma r)}}{r},
\end{align}
where $r$ is depth and $\gamma$ is the absorption.

\paragraph{Noise Modeling}
The noise in system 1 needs to be investigated in a detailed manner. However, a superficial noise model is available in literature \cite{Tonolini:SIGCOM:2018}, where the noise power is given as:
\begin{align}
    P_\text{N}^{\text{TARF}}(\omega)=\frac{\rho v \omega^2}{a},
\end{align}
where $a$ is a real positive constant which depends on the distance attenuation, noise floor and transmit power.

\subsubsection{System 2: Optical based A-W:} 
The channel of such systems can be characterized through transmittance (can be thought as inverse of path-loss). Following \cite{guo2018channel}, we discuss the transmittance through three regions below. \\
\paragraph{Air part} The transmittance in air medium can be expressed as:
\begin{align}
    T_{\text{Air}}= C_0^\text{T} \exp{\left(-\biggl[\frac{{r}/{r_0}}{S(\frac{2}{R_{\text{eff}}(\phi-\psi_0)})} \biggl]^{S`(\frac{2}{R_{\text{eff}}(\phi-\psi_0)})}\right)},
\end{align}
where $r$ is the beam deflection distance, $r_0$ is the aperture radius $C_0^\text{T}$ is the maximum transmission coefficient at $r=0$, $R_{\text{eff}}(.)$ is the effective spot radius, $S(.)$ is scale function and $S`(.)$ is shape function. \\
\paragraph{Surface part} The transmittance of the surface is given as:
\begin{align}
    T_{\text{surface}}=\eta_{\text{deflection}}T_{\text{refraction}}
\end{align}
where $\eta_{\text{deflection}}$ denotes efficiency due to deflection of light because of surface roughness, its computation is given in \cite{guo2018channel},  and $T_{\text{refraction}}$ is the total Fresnel transmittance given below.
\begin{align}
    T_{\text{refraction}}= \frac{T_\perp +T_\parallel}{2}
\end{align}
where $T_\perp$  is the perpendicular and $T_\parallel$ is the parallel components of Fresnel transmittance, which can be computed as
\begin{align}
    T_\perp= 1-\frac{sin^2(\theta_\text{i}-\theta_\text{r})}{sin^2(\theta_\text{i}+\theta_\text{r})} \\
    T_\parallel=1-\frac{tan^2(\theta_\text{i}-\theta_\text{r})}{tan^2(\theta_\text{i}+\theta_\text{r})}, \nonumber
\end{align}
where $\theta_\text{i}$ and $\theta_\text{r}$ are incident and refraction angles respectively.
\\
\paragraph{Underwater part} The transmittance in water $T_\text{W}$ can be expressed as
\begin{align}
    T_\text{W}=10^{-K/10},
\end{align}
where K is the total attenuation coefficient.
\\
\paragraph{Noise Modeling} The total noise power is given as
\begin{align}
    P_{\text{TN}}^{\text{OA-W}}=\text{ENL}+T\frac{P_{\text{N}}^\text{A}\exp{(-K\delta)}+P_\text{N}^\text{W}}{hf_\text{n}}
\end{align}
Where $\text{ENL}$ is excess noise limit, $T$ is the sampling period of detector, $K$ is the attenuation coefficient, $\delta$ is the depth, $h$ is Plank constant, $f_\text{n}$ is the noise frequency $P_{\text{N}}^\text{A}$ is nose power in air  and  $P_\text{N}^\text{W}$ is noise power in water/sea. The can be expressed as
\begin{align}
    P_{\text{N}}^\text{A}=B_\text{s}\text{FOV}\pi r^2 \Delta f,
\end{align}

where $B_\text{s}$ is the sky brightness, $r$ is the radius of virtual telescope  and $\Delta f$ is the filter bandwidth and $P_\text{N}^\text{W}$ can be expressed as
\begin{align}
P_\text{N}^\text{W}= R_{\text{solar}} \text{FOV} \pi r^2 \Delta f,
\end{align}
where $R_{\text{solar}}$ is the solar radiance given in Eq. \ref{eq:R_sol}. \\

\subsubsection{System 3: MI based A-W:}
The channel characteristics of MI based A-W wireless communication in both air and water parts are similar since the magnetic permeability of both air and water are the same \cite{li2019survey}. For the air part, the path-loss expression is given in Eq. \ref{eq:pl_MI}, while for water part, the path-loss expression is given in Eq. \ref{eq:pl_MI_total}, which includes the additional path-loss due to eddy current (see Eq. \ref{eq;pl_EC}) \cite{domingo2012magnetic}.

The noise modeling for MI based A-W communication is similar to that discussed in section \ref{subsec:MI_UWCNs} for MI based underwater communication. 

\section{Security of underwater and A-W WCNs}
\label{sec:4}

In sections \ref{sec:2} and \ref{sec:3}, We described the system models, technological enablers, unique challenges of underwater medium, and various system configurations. This should have intrigued the curious reader to anticipate and imagine the various different kinds of attacks that could be launched on underwater and A-W WCNs. For example, one could see that the optical-based underwater and A-W channels (being highly directive in nature) are least prone to passive eavesdropping attacks by the malicious nodes nearby, especially when compared to their RF and acoustic counterparts. One could also verify that launching a jamming attack on an optical-based system is difficult because the jammer will need to move to the field-of-view of the optical receiver, where the optical jammer will very likely be detected. Similarly, it is not difficult to see that launching an impersonation (stealth) attack on RF-based underwater and A-W WCNs is relatively straightforward; mainly because typically it is much easier to localize a malicious acoustic or optical source (being audible or visible, intermittently).

\begin{figure*}[htb!]
\centering
\includegraphics[width=1\textwidth]{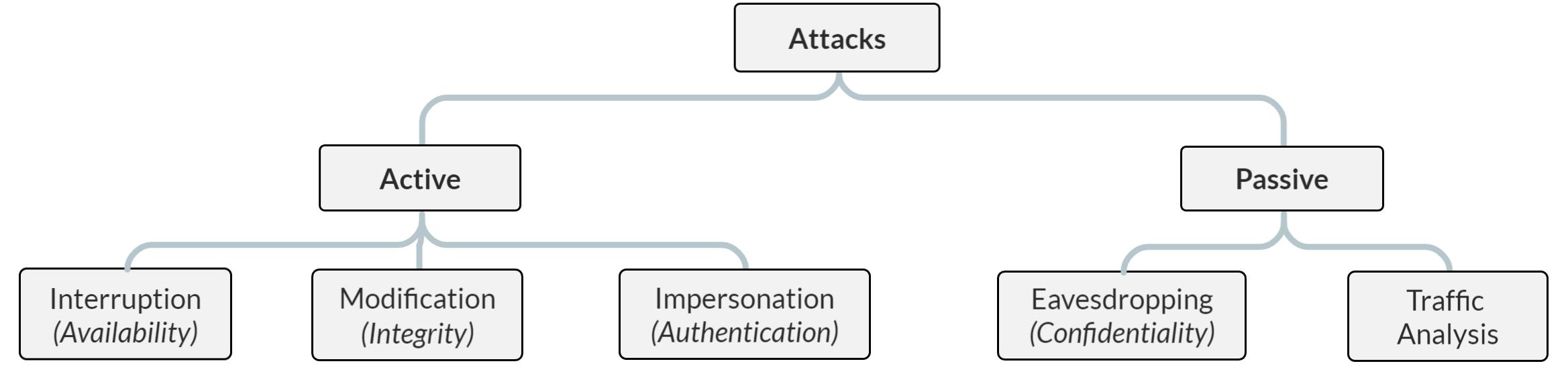}
\caption{Broad classification of malicious attacks}
\centering
\label{fig:Attack}
\end{figure*}

This brings us to the main contribution of this paper where we provide a thorough and up-to-date survey of the relevant papers in the field addressing the essential properties of information security as shown in Table \ref{tb:1:2}. However, before we delve into the security of underwater and A-W communications, it is imperative to provide a quick background on the domain of information security.
\begin{table*}[htb!]
\centering

\begin{adjustbox}{{width=1\textwidth}}
\small
\begin{tabular}{ |l|c|p{4cm}|p{3cm}|c|c|c|}
 \hline

 \hline
 
\bf{Security Property}  &  \multicolumn{4}{c|}{\bf{Underwater Wireless Communication}} & \multicolumn{2}{c|}{\bf{Air-Water Wireless Communication}}    \\

 &  RF & Acoustic & Optical& MI & Indirect &Direct  \\ \hline
Confidentiality & - & \cite{Liu:ICSP:2008}  \cite{DiniCC:ISCC:2011}  \cite{ Spaccini:OCEANS:2015} \cite{Dai:Sensors:2016} \cite{Huang:SJ:2016} \cite{Huang:TWC:2016}  \cite{Xu:TMC:2018}  \cite{pelekanakis2019towards} \cite{pelekanakis2019robust} \cite{Bagali:IJECE:2020} \cite{bagali2020efficient} \cite{chiariotti2020underwater} \cite{liang2020cs}  \cite{waqas:ett:2020} \cite{Ming:WCL:2020} \cite{Signori:IoTJ:2020} \cite{ozmen2020impact} \cite{Li:Senors:2020} \cite{signori2021geometry} \cite{Yu:ICAICA:2021} \cite{Jain:CCPE:2021} \cite{pelekanakis2021physical}
\cite{sklivanitis2021physical} &\cite{Kong:OE:17}
\cite{Verma@2021}
\cite{Shaboy:OE:18}
\cite{Du:OE:21} \cite{shi2015channel} \cite{peng2019performance} \cite{gariano2019theoretical} \cite{mao2020monte} \cite{wang2020improving} \cite{xiang2021improving} \cite{zuo2021security} & -&\cite{Illi:Access:2018} \cite{  Lou:ICL:2021} \cite{Illi:ITSC:2021} \cite{Badrudduza:Access:2021} \cite{lou:ArXiv:2020} & \cite{xu2018performance} \cite{guo2018channel} \cite{xie2018security} \cite{peng2022satellite} \\
\hline
Integrity  & - & \cite{ DiniND:ISCC:2011} \cite{ Spaccini:OCEANS:2015}  \cite{Diamant:TWC:2019}   \cite{campagnaro2020replay} \cite{Yu:ICAICA:2021} & - & -& - & - \\
\hline
Authentication  & - & \cite{Liu:ICSP:2008}
\cite{Zhang:INFOCOM:2010}
\cite{DiniCC:ISCC:2011} \cite{ DiniND:ISCC:2011}
\cite{ Souza:ISCC:2013} \cite{Li:SoftCOM:2013}
\cite{li2015spoofing}
\cite{ Spaccini:OCEANS:2015}
\cite{Huang:TWC:2016}
 \cite{Xiujuan:IJDSN:2017}
 \cite{liu2017secure}
\cite{Bhar:IJCS:2017}
\cite{Xu:TMC:2018}
\cite{xiao:ICLJamming:2018}
\cite{Aman:Access:2018}
  \cite{Xiao:ICL:2018}
\cite{Diamant:TWC:2019}
\cite{pelekanakis2019towards}
\cite{pelekanakis2019robust} 
 \cite{liang2020cs}
 \cite{Saeed:Access:2020}
\cite{khalid2020physical} 
\cite{khalid2020node}
 \cite{das2020anomaly}
\cite{Ming:WCL:2020}
\cite{Pan:TWC:2021}
\cite{Yu:ICAICA:2021}
\cite{Zala:ICCCNT:2021}
\cite{Jain:CCPE:2021}
\cite{Bragagnolo:COMCAS:2021}
\cite{pelekanakis2021physical}
\cite{sklivanitis2021physical}
\cite{Islam:ICEEICT:2021}
\cite{Zhao:ISJ:2022}
\cite{alharbi2022securing} & \cite{shi2015channel} \cite{peng2019performance} \cite{gariano2019theoretical} \cite{mao2020monte} \cite{wang2020improving} \cite{xiang2021improving} \cite{zuo2021security} & -& - & \cite{xu2018performance} \cite{guo2018channel} \cite{xie2018security} \cite{peng2022satellite}
\\
\hline
Availability & -&\cite{Zhang:INFOCOM:2010} \cite{Goetz:IWUWN:2011} \cite{ DiniND:ISCC:2011} \cite{Misra@2012} \cite{Zuba:SCN:2015} \cite{Xiao:Globecom:2015}  \cite{liu2017secure}
\cite{Bhar:IJCS:2017} \cite{xiao:ICLJamming:2018} \cite{bagali2019efficient} \cite{Bagali:IJECE:2020}
 \cite{bagali2020efficient} \cite{Saeed:Access:2020} \cite{Saeed:Access:2020} \cite{Li:Senors:2020} \cite{signori2021geometry} \cite{Zala:ICCCNT:2021}  \cite{alharbi2022securing}&-& - &- &- \\ \hline
\end{tabular}
\end{adjustbox}
\caption{Current state-of-the-art in underwater and air-water wireless communication: each work has been mapped to one of the four main properties of information security.}
\label{tb:1:2}
\end{table*}

\subsection{A Primer on Information Security}
Information security aims to protect information exchanged between legitimate parties. Any communication system with the provisioning of information security should ensure that the following properties (also known as four fundamental properties of information security) are preserved:
\begin{itemize}
 
    \item \textit{Confidentiality}: ensures secrecy of transmitted information.
    \item \textit{Authenticity}: ensures legitimacy of exchanging parties.
    \item \textit{Integrity}: ensures that information exchanged between legitimate nodes have not been modified.
    \item \textit{Availability}: ensures the availability of the system for legitimate nodes.
 
\end{itemize}

To preserve these properties, two broad approaches are adopted: traditional cryptography and Physical Layer Security (PLS), which we briefly discuss next.

\subsubsection{Traditional Cryptography} Typically, to establish secure communication, some cryptographic measures need to be achieved, where legitimate nodes jointly solve a complex mathematical problem (e.g., elliptic curves) to establish shared secrets, that can then be used to preserve confidentiality, authentication and integrity. Crypto-based systems can be divided into two main classes, symmetric and asymmetric key cryptography.

\begin{itemize}
    \item Symmetric key: also known as private-key cryptography is an approach for encrypting and decrypting information with a pre-shared secret key. The most popular symmetric key algorithm is Advanced Encryption Standard (AES).
    \item Asymmetric key: also known as public-key cryptography, is an approach for encrypting and decrypting information using two keys, a public key (known to everyone) and a private (known to the owner only). Messages encrypted by one key can only be decrypted by the other. The most popular asymmetric key algorithms are RSA \cite{rivest1978method} and DH \cite{diffie1976new}. 
\end{itemize}

However, recent advances in computational resources and quantum computing, may potentially jeopardize these cyrpto-based solutions \cite{shor1994algorithms,arbaugh:IWC:2002,gidney:arXiv:2019,duan2020proof}. This has motivated the community to explore alternative approaches, a popular one is: Physical Layer Security.

 
\subsubsection{Physical Layer Security (PLS)}
PLS is increasingly becoming an attractive area of research due to its promising performances in many wireless communication systems \cite{shiu:IWC:2011,bloch2011physical,Zhou:Book:2013,Waqas:WCNC:2016,Ammar:VTC:2017,Waqas:ett:2018,Waqas:VTC:2019,waqas:UCET:2019,waqas:UCET:2020,waqas:Sensors:2021}. PLS exploits the random nature of the physical layer for security purposes, which not only enhances security, but also greatly improve efficiency. In wireless communication, randomness is introduced due to two main sources: wireless channel and hardware. The randomness in the wireless channel is introduced due to random noise and random multi-paths arrival, which forcefully randomizes some processes such as Channel Impulse Response (CIR), Channel Frequency Response (CFR), and Received Signal Strength (RSS). On the other hand, randomness in hardware is introduced due to uncontrollable impurities added during the manufacturing process. 

\begin{figure*}[htb!]
\centering
\includegraphics[width=1\textwidth]{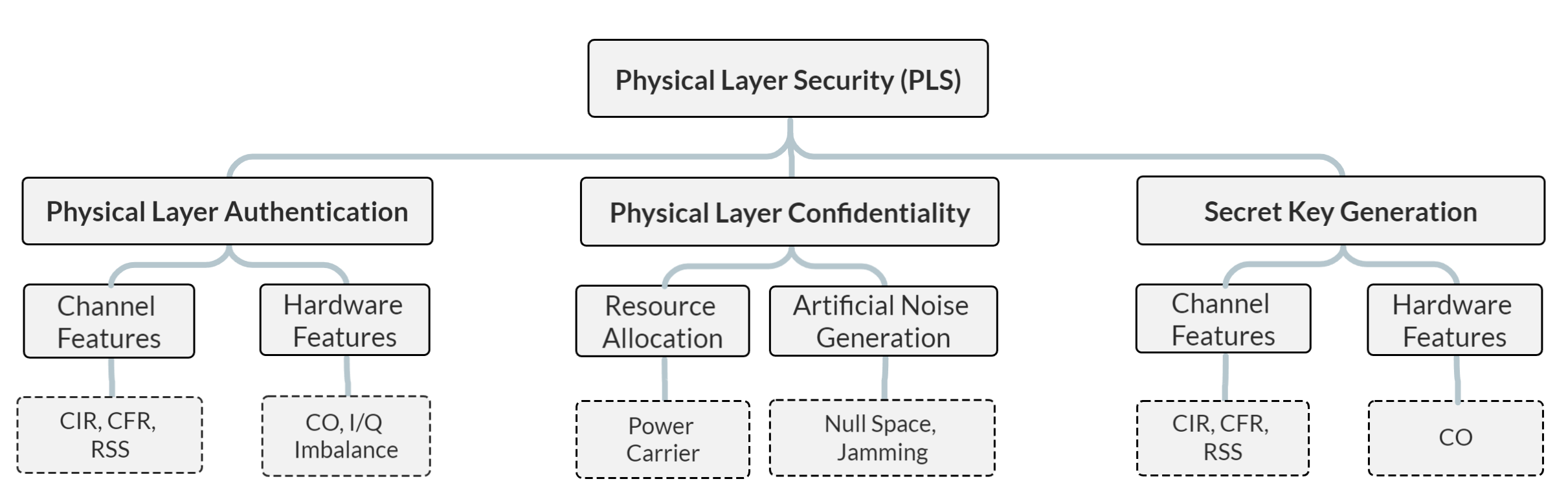}
\caption{Security properties preserved by PLS. The Figure also lists examples of features that can be used to provide these properties.}
\centering
\label{fig:pls}
\end{figure*}

Like traditional cryptography, PLS attempts to preserve a number of important security properties as illustrated in fig. \ref{fig:pls}. 

\begin{figure}
\centering
\includegraphics[width=0.5\textwidth]{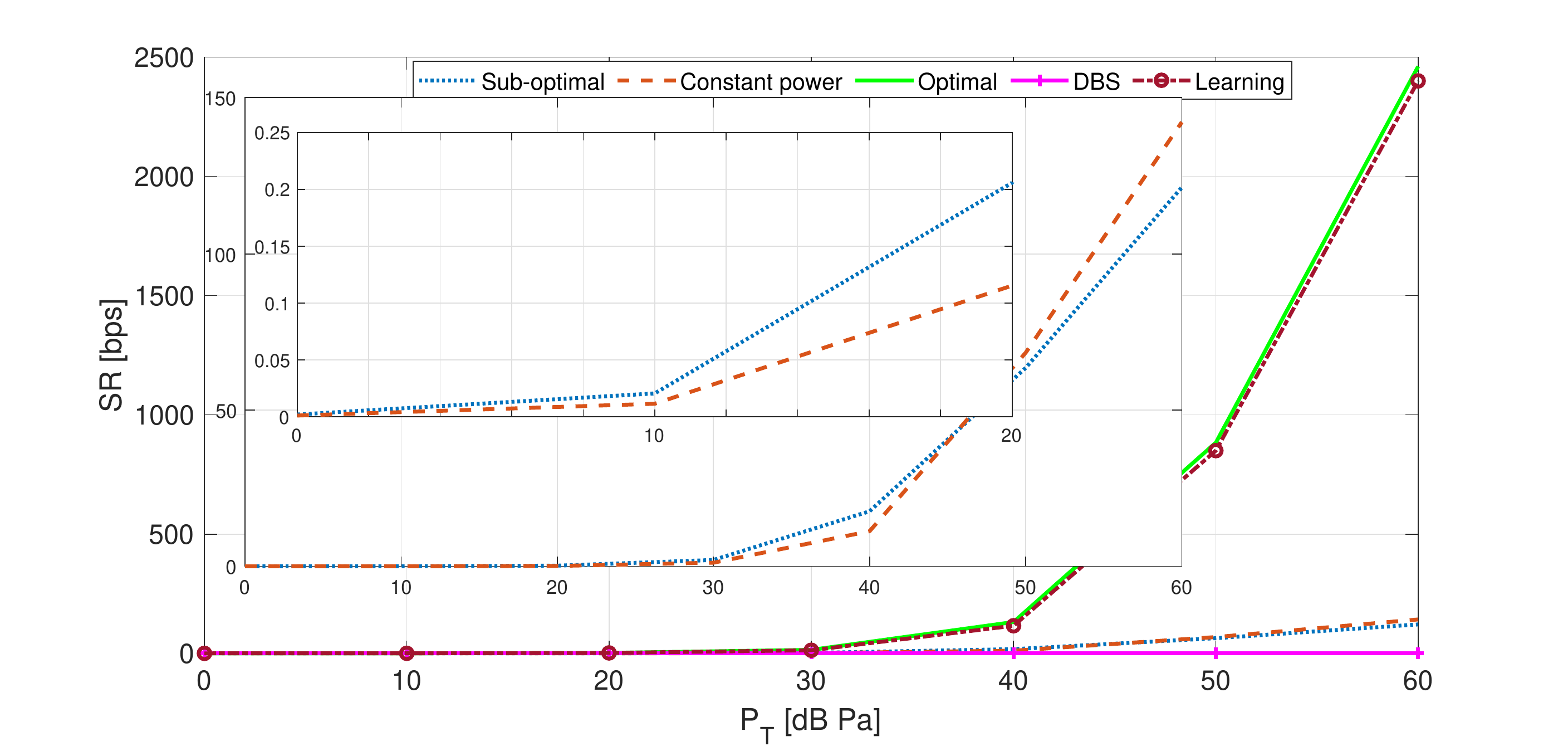}
\caption{Secrecy rate (SR) vs total power budget $P_t$ of underwater acoustic communication system considered in \cite{waqas:ett:2020}. The performance of proposed optimal scheme is compared against other proposed schemes (i.e., learning and sub-optimal) and baseline schemes (i.e., Depth Based Selection (DBS) and constant power). }
\centering
\label{fig:serecy}
\end{figure}

\noindent\textit{Physical layer confidentiality}  provides security to information in-transit. In other words, this mechanism tackles the eavesdropping attacks where the eavesdropper is in listening mode. The basis of this mechanism was first introduced by Wyner in his seminal work \cite{Wyner:BSTJ:1975}. The secrecy rate and the probability of leakage or secrecy outage (which depends on information-theoretic bounds) are the two main performance metrics for physical layer confidentiality. The secrecy rate can be enhanced through optimal resource allocation \cite{Waqas:WCNC:2016} (such as depicted in Fig. \ref{fig:serecy}, taken from one of our previous works \cite{waqas:ett:2020}) and/or artificial noise generation \cite{ANPLS:WCSP:2016}, where noisy signals are generated in the null space of the legitimate users. Through similar procedures, secrecy outages can be minimized. Resources such as carrier power, carrier selection, relay selection, and user selection are exploited for physical layer confidentiality \cite{bloch2011physical}, \cite{Zhou:Book:2013}.

\begin{figure}
\centering
\includegraphics[width=0.5\textwidth]{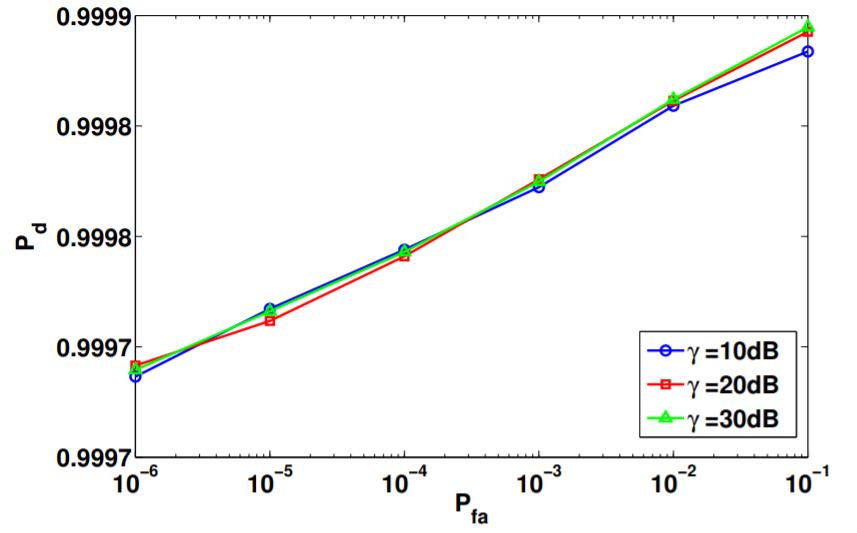}
\caption{ROC: probability of detection $P_d$ against probablity of false alarm $P_{fa}$ for various values of SNR $\gamma$ \cite{Mahboob:Globecom:2014}.}
\centering
\label{fig:ROC}
\end{figure}

\noindent\textit{Physical layer authentication} is a systematic procedure that verifies the legitimacy of the transmitter node based on the characteristics of the physical layer, such as device fingerprints or features. Physical layer authentication generally involves two steps: \textit{feature estimation} and \textit{testing} \cite{Bai:PLA:2020}. This mechanism requires a feature that is random in nature and independent for distinct transmitters. The reported features can be classified into medium/channel-based features and hardware-based features (or RF-fingerprints). CIR \cite{Ammar:VTC:2017}, CFR \cite{CFR1:TWC:2008},  RSS \cite{RSS:TPDS:2013} are examples of channel-based features. In-phase/Quadrature (I/Q) imbalance \cite{IQ:ICC:2014} and carrier offsets (CO) \cite{Mahboob:VTC:2017}, \cite{Mahboob:Globecom:2014} are examples of hardware features. Error probabilities (false alarm and miss detection which also produce Receiver Operating Characteristics (ROC) curves (depicted in Fig. \ref{fig:ROC}, taken from one of our previous works \cite{Mahboob:Globecom:2014} )), Kullback-Leibler Divergence (KLD) \cite{Mahboob:Globecom:2014} \cite{CFR2:TWC:2012} and Jensen-Shannon Divergence (JSD) \cite{JSD:ICNNA:2017} are key performance metrics to evaluate physical layer authentication mechanisms.

\begin{figure}
\centering
\includegraphics[width=0.5\textwidth]{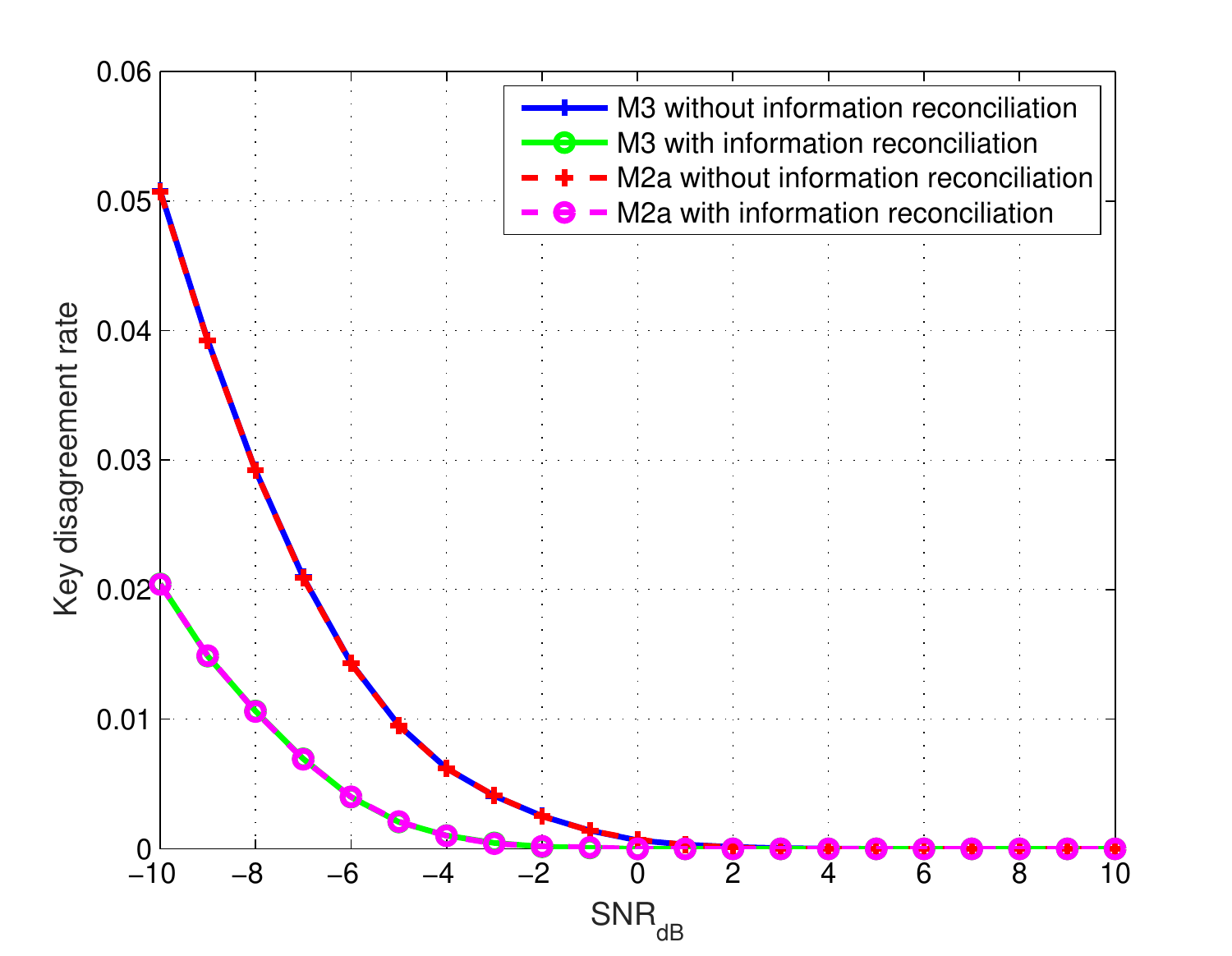}
\caption{Key disagreement rate against SNR (in dB) for two proposed CFO models (i.e., M2a, where CFO is time-varying and memory-full and M3, where CFO is time varying and memoryless) \cite{Waqas:VTC:2019}.}
\centering
\label{fig:kdr}
\end{figure}

\noindent\textit{Shared secret key generation} is a process where a pair of legitimate users extract a secret key from commonly shared time-varying sources (features/attributes of the physical layer). This mechanism aims to avoid the dependency on pre-shared secret keys. This mechanism requires a reciprocal time-varying sources or features among the legitimate users to extract secret keys.  \textit{Feature estimation}, \textit{quantization},  \textit{key reconciliation} and \textit{privacy amplification} are standard steps used to extract the keys. Mutual information, key generation rate, key disagreement rate (illustrated in Fig. \ref{fig:kdr}, taken from one of our previous works \cite{Waqas:VTC:2019}) bit-match rate, burst-match rate, and decipher probability are key performance metrics used to evaluate this mechanism's performance. RSS, CIR \cite{SKG:SCN:2014}, CFR \cite{CFRSKG:ICCN:2014}, CO \cite{Waqas:VTC:2019} are examples of features that can be used to generate a secret key. The generated secret keys can then be used to provide both confidentiality (by using any symmetric cryptography approaches) and authentication (by using generated keys as transmit device fingerprint \cite{MT:ICRATE:2014}) in communication systems.

%
%

\subsubsection{Attack Types}
Common malicious attacks jeopardizing the above security properties (i.e., confidentiality, authentication, integrity, availability) can be broadly classified as active and passive attacks. In active attacks, a malicious node transmits malicious signals to obtain illegal access or disrupt ongoing communication. In passive attacks, a malicious node does not transmit any signals but listens passively to the legitimate nodes' ongoing communication. A broad classification of attacks, and what security properties they compromise, is illustrated in Fig. \ref{fig:Attack}. Examples of these attacks, include: 
\begin{itemize}
    \item Spoofing: also known as impersonation, where a malicious node claims the identity of a legitimate node.
    \item Replay: in this attack, a malicious node re-uses intercepted legitimate messages.
    \item Sybil: in this attack, a malicious node claims the identities of many legitimate nodes.
    \item TSA: in this attack, a malicious node tampers with legitimate nodes' time synchronization, which is a fundamental requirement for localization purpose. The goal of the malicious node is to convince some nodes that their neighbors' clocks are at a different time than they actually are.
    \item Wormhole: two or more malicious nodes jointly launch this attack. At one point of the network, the attacker receives packets and tunnel them to  another point in the network and then replay them from  that point. This attack typically disrupts the routing procedure of the network. 
    \item Blackhole: in this attack, a malicious node advertises convincing parameters to the legitimate nodes in order to become a forwarding node in their routing path. The goal of  the attacker is to create a hole in the network when incoming packets can be dropped.
    \item Jamming: in this attack, a malicious node bombards legitimate nodes with an excessive amount of signals potentially making them unavailable to process anything else. Jamming is typically either active jamming or reactive jamming. In former, a jammer node produces jamming signals continuously while in later, a jammer transmits whenever communication between legitimates parties is detected.
    \item Eavesdropping: in this attack, a malicious node is interested in the communication between legitimate parties and tries to listen and decode intercepted traffic.
    \item DoS: in this attack, a malicious node attempts to prevent the legitimate nodes from accessing network resources. Typically, DoS is achieved through jamming and flooding. Therefore, Jamming, Wormhole, blackhole are considered as DoS attacks. 
\end{itemize}


\subsection{Security of Underwater Wireless Communication}
\begin{figure}
\centering
\includegraphics[width=0.5\textwidth]{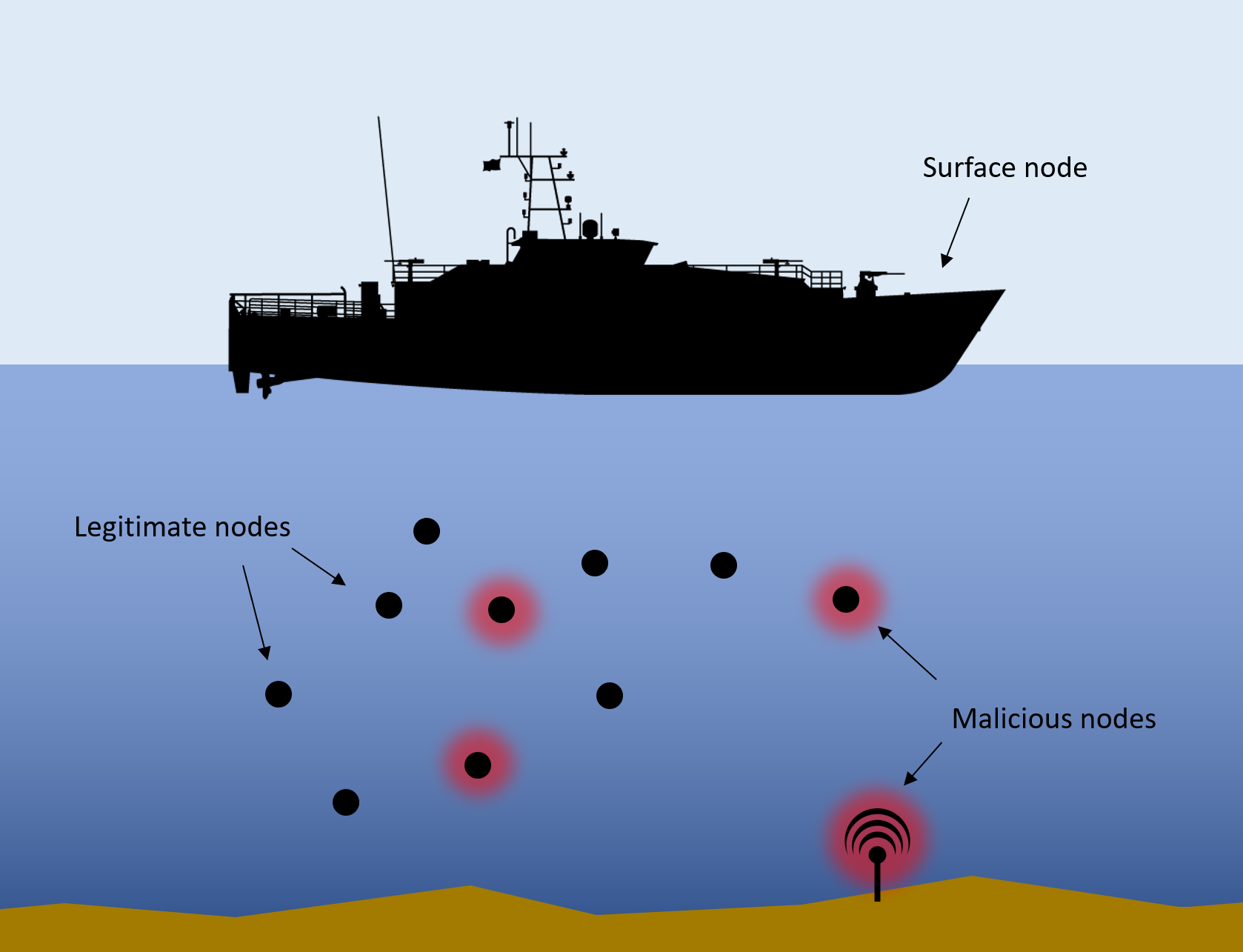}
\caption{Illustration of a scenario where malicious nodes present in underwater launch active and passive attacks. The malicious seabed node is depicted as an active malicious node, while the other is a passive malicious node.}
\centering
\label{fig:singel-medium}
\end{figure}

Fig. \ref{fig:singel-medium} illustrates a scenario where adversaries present in the close vicinity of legitimate nodes perform underwater malicious attacks. The attacks can be active or passive. 

Recall from Section-\ref{sec:2} that underwater wireless communication can be accomplished using RF, acoustic and optical communication technologies. Hence, we now discuss the most popular attacks on each underwater communication technology.

\subsubsection{Security of RF-based Underwater WCN}
To the best of our knowledge, there are no recent advances on RF-based underwater WCN due to their shortcomings discussed in section \ref{subsec:RF}. Hence, we found no work studying attacks on such systems. 



\subsubsection{Security of Acoustic-based Underwater WCN}
Underwater acoustic communication became an active research area in the last few decades, which resulted in quantitative and qualitative research on the security of underwater acoustic communication. We summarize the state-of-the-art security in underwater acoustic communication in terms of confidentiality, authentication, and integrity, in Tables \ref{tb:3},  \ref{tb:3B} \& \ref{tb:3C}. Table \ref{tb:3} summarizes the relevant works published before 2019; Table \ref{tb:3B} summarizes relevant works published during the period 2019-20; while Table \ref{tb:3C} summarizes relevant works published from 2021 to date. 
The most common scenarios assumed for underwater communications are multi underwater nodes (more than two legitimate nodes and one or more malicious node) and three underwater nodes (i.e., two legitimate nodes and a malicious node). Furthermore, single carrier, multi-carrier (i.e., Orthogonal Frequency Division Multiplexing (OFDM)), Line-of-Sight (LoS), and non-LoS underwater acoustic communications are exploited for security purposes. In short, various system configurations are judged for their strength and weaknesses against attacks. Some configurations will do better than others, for each kind of attack. The counter-mechanism strategies borrow tools from cryptography, optimization theory, game theory, control theory, probability theory and machine learning to equivocate/confuse a passive malicious node, or to partially recover data amid an active jamming or spoofing attack by a malicious node. 


As shown in Tables \ref{tb:3}, \ref{tb:3B} \& \ref{tb:3C}, the most common attacks considered in underwater acoustic communication are spoofing (also known as impersonation), eavesdropping, and Denial of Service (DoS). Variants of these attacks are also sometimes considered, such as the replay, sybil, masquerade, wormhole, Time Synchronization Attack (TSA), Blackhole and time delay attacks, which are variants of the spoofing attacks. Jamming, Wormhole and Blackhole are also considered DoS attacks \cite{sharma2018survey}. Tables \ref{tb:3}, \ref{tb:3B} \& \ref{tb:3C} also provide details about some countermeasures that have been shown to resist these attacks.

\subsubsection{Security of Optical-based Underwater WCN}
\label{sssec:security:UWOC}
Optical-based underwater WCN is generally considered a relatively more secure communication approach (compared to its acoustic and RF-based counterparts) due to the high directivity of the light beam~\cite{Illi:ITSC:2021}. However, an eavesdropper located in close vicinity to the transmitter and receiver might disturb the information transmission by partially blocking the transmitting light-beam~\cite{6685978}. Indeed, the eavesdropper might require a sophisticated device such as a beam splitter as the laser beam near the transmitter node becomes narrow. Therefore, the authors in \cite{Kong:OE:17} experimentally illustrated some vulnerabilities in underwater wireless optical communication. They use reflecting mirrors between communication paths to measure any leakage towards the eavesdropper~\cite{Verma@2021}. Similarly, authors in \cite{Shaboy:OE:18} used diffraction grating to tap the optical channel. Recently, the experimental work in \cite{Du:OE:21} studied the chaotic encryption in 50m/5Gbps using 450 nm laser to provide confidentiality. Furthermore, the laser beam experiences divergence due to the \emph{scattering} phenomenon~\cite{7968296}. In principle, the light wave strikes an aerosol (i.e. dust particles or gas molecules) and scatters into several light beams in different directions~\cite{Badrudduza:Access:2021}. Therefore, it is possible that an eavesdropper located near the receiver can capture the scattered photons and hence intercepts the legitimate information transmission.

On the other hand, recently,  Quantum Key Distribution (QKD) has been explored theoretically based on optical-based underwater WCN \cite{peng2019performance,mao2020monte,wang2020improving,xiang2021improving,zuo2021security}. QKD is a promising mechanism where secret keys are generated based quantum mechanical properties, such as entanglement \cite{shi2015channel,gariano2019theoretical,peng2019performance,
mao2020monte,wang2020improving,xiang2021improving,zuo2021security}; the no-cloning theorem \footnote{In Quantum mechanics, the no-cloning theorem states that it is not possible to make a copy of a Quantum state as copying it will entail observing it, and observing it will inevitably destroy it.} \cite{wootters1982single} then guarantees that these keys cannot be intercepted. QKD is divided into Discrete Variable QKD (DV-QKD) \cite{Gisin:RMP:2002, lo2014secure,pirandola2015high} and Continuous Variable QKD (CV-QKD) \cite{RevModPhys77513, PhysRevA81062343,jouguet2013experimental,gehring2015implementation}, where the latter has recently been preferred due to easy implementation and higher channel capacity \cite{peng2019performance,mao2020monte,wang2020improving,xiang2021improving,zuo2021security}.

\begin{table*}[htbp]
\centering
\begin{adjustbox}{{width=1\textwidth}}
\tiny
\begin{tabular}{ |p{1cm}|p{1.5cm}|p{1.5cm}| p{3cm}| p{2cm}|}
 \hline
\bf{Ref.} &\bf{System Architecture}  & \bf{Attacks considered} & \bf{Countermeasures proposed} &  \bf{Security Properties addressed}\\
 \hline
 \cite{Liu:ICSP:2008} (2008) &Three nodes (two legitimate and one malicious nodes) Underwater Acoustic Sensor Network (UWASN)   &   Spoofing and Eavesdropping & Shared secret key generation. Feature:Envelope of tone signal.     Robust Secure Fuzzy Information
Reconciliation (RSFIR)  &   Confidentiality and Authentication \\
  \hline
  \cite{Zhang:INFOCOM:2010} (2010)& Multi-nodes single-hop UWASN & Wormhole &Direction of arrival (DoA) for each node is estimated to make the network resilient to wormhole attack during neighbour discovery for suit of protocols&Authentication and Availability\\ \hline
  \cite{Goetz:IWUWN:2011} (2011) &Multi-nodes multi-hop UWASN & DoS (Jamming)& Multi-path routing & Availability \\ \hline
 
\cite{DiniCC:ISCC:2011} (2011)   & Underwater acoustic network of AUVs   & Spoofing {and} Eavesdropping   & Crypto-based suit &  Confidentiality, Authentication and Integrity \\ \hline
\cite{ DiniND:ISCC:2011} (2011)   & Underwater acoustic network of AUVs    & Spoofing {and} DoS   & Crypto-based scheme: Secure FlOOD (SeFLOOD), where FLOOD is a network discovery protocol  &  Authentication, Integrity and Availability \\\hline
\cite{Misra@2012} (2012) &Multi-nodes multi-hop UWASN &Jamming & Underwater Jamming Detection Protocol (UWJDP): packet send ratio, packet delivery ratio and energy consumption amount are used.   &Availability \\ \hline
\cite{ Souza:ISCC:2013} (2013)  &Multi-nodes multi-hop UWASN   & Spoofing   & Crypto-based scheme: Digital signatures generated using three schemes-ECDSA, BLS and ZSS. &  Authentication \\\hline
\cite{Li:SoftCOM:2013} (2013) &Multi-nodes multi-hop UWASN & Sybil & Received packet and time rates, disposed and identification packet rates are used.  & Authentication\\ \hline
\cite{Zuba:SCN:2015} (2015) & Two legitimate nodes UWASN with multiple jammers& Jamming & No countermeasures are provided. Detailed experimental results are demonstrated. The experiments are conducted in Mansfield Hollow Lake, CT, USA & Availability\\ \hline
\cite{Xiao:Globecom:2015} (2015) & CSMA-based multiple-nodes single-hop UWASN &  Jamming & Static and dynamic jamming games. Nash equilibrium for static game and re-enforcement learning based solution for dynamic game are proposed &Availability \\ \hline 
\cite{li2015spoofing} (2015) &Multi-nodes single-hop UWASN & Spoofing&Game theoretic approach: Nash Equilibrium is derived for zero-sum game between a spoofer and sensors & Authentication \\ \hline
\cite{ Spaccini:OCEANS:2015} (2015)  &Multi-nodes single-hop UWASN   & Spoofing  and Eavesdropping  & Crypto-based schemes:  symmetric and asymmetric keys based encryption and authentication &  Confidentiality, Authentication and Integrity \\
  \hline
  \cite{Dai:Sensors:2016} (2016)  &Multi-nodes UWASN  &  Eavesdropping  & No countermeasures: probability of eavesdropping is calculated using stochastic geometry &  Confidentiality \\
\hline 
\cite{Huang:SJ:2016} (2016) &OFDM-based three-nodes UWASN &  Eavesdropping  & Jamming is used to enhance the secrecy rate &  Confidentiality \\
\hline 
\cite{Huang:TWC:2016} (2016) & OFDM based three nodes UWASN &  Spoofing and Eavesdropping & Feature: Amplitudes of Channel Frequency Response (CFR).   Experimentation in  Mansfield Hollow Lake, Connecticut, USA. Amplitudes of CFR is estimated and then quantized using CDF based quantization. BCH codes are used for key reconciliation   & Confidentiality and Authentication \\
 \hline
 \cite{Xiujuan:IJDSN:2017} (2017) & Multi-nodes multi-hop UWASN & Spoofing (Forgery) & Secure routing: two-way authentication through use of digital signatures & Authentication \\\hline
 \cite{liu2017secure} (2017) & Multi-nodes multi-hop UWASN &Wormhole & Secure localization scheme & Authentication and Availability \\ \hline
 \cite{Bhar:IJCS:2017} (2017) & Multi-nodes multi-hop UWASN &Wormhole & Agent-based Secure routing scheme & Authentication and Availability \\ \hline
\cite{Xu:TMC:2018} (2018)& OFDM-based UWASN & Spoofing and Eavesdropping &Shared secret key generation. Feature: Phases of CFR.  Deterministic quantization approach, called Virtual Phase Shift (VPN) is used for key generation &  Confidentiality and Authentication\\
 \hline
 
   \cite{Aman:Access:2018} (2018) &  Multi-nodes single-hop UWASN   & Spoofing   & Physical layer security: A novel three features based authentication&  Authentication \\
  \hline
  \cite{Xiao:ICL:2018} (2018) & Three nodes UWASN &Spoofing &Reinforcement-learning-based and deep-reinforcement-learning-based authentication mechanisms: average power and delay in channel power delay profile are used as features & Authentication \\ \hline
  \cite{xiao:ICLJamming:2018} (2018) &Two legitimate and a malicious acoustic nodes &Jamming &Q-learning based scheme: Transmit power and node mobility is used to mitigate jamming & Availability\\ \hline
 \end{tabular}
 \end{adjustbox}
 \caption{Summary of the reported security work in  underwater acoustic communication systems (Upto 2019). }
 \label{tb:3}
\end{table*}

\begin{table*}[htbp]
\centering
\begin{adjustbox}{{width=1\textwidth}}
\tiny
\begin{tabular}{ |p{1cm}|p{1.5cm}|p{1.5cm}| p{3cm}| p{2cm}|}
 \hline
\bf{Ref.} &\bf{System Architecture}  & \bf{Attacks considered} & \bf{Countermeasures proposed} &  \bf{Security properties addressed}\\
 \hline
    \cite{bagali2019efficient} (2019) & Multi-nodes UWASN with single jammer&Jamming& Channel access model based on cross-layer design & Availability\\ \hline
   
  \cite{Diamant:TWC:2019} (2019)  & Multi-nodes single-hop UWASN   & Spoofing    & Cooperative authentication based on statics of estimated channel characteristics &  Authentication and Integrity \\
  \hline
  \cite{pelekanakis2019towards} (2019) & Three-nodes UWASN & Spoofing and Eavesdropping & CIR features: L$_2$ /L$_0$ norm, a smooth sparseness measure and the root-mean-square delay spread.  Thousands of channel-probe signals  are used to extract the CIR features  &  Confidentiality and Authentication\\
 \hline
  \cite{pelekanakis2019robust} (2019) &Three-nodes UWASN & Spoofing and Eavesdropping & CIR features: L$_2$ /L$_0$ norm, channel sparseness, channel energy, root-mean-square delay spread, mamximum delay spread and sum of successive delays.  Thousands of channel-probe signals  are used to extract the CIR features  &  Confidentiality and Authentication\\ \hline
  \cite{Bagali:IJECE:2020} (2020)&Multi-nodes UWASN with single jammer& Jamming & Resource allocation scheme&Confidentiality and Availability\\ \hline
  \cite{bagali2020efficient} (2020) &Multi-nodes UWASN with multiple jammers &Jamming &   Resource allocation scheme & Confidentiality and Availability\\ \hline
  \cite{chiariotti2020underwater} (2020) & Two legitimate nodes and a jammer  & Jamming &Game theory: optimal strategies are derived for both nodes (defender and attacker) to estimate attacker's location. Bayesian Nash Equilibrium (BNE) is derived for considered Bayesian Jamming game &Confidentiality and Availability \\ \hline
 \cite{liang2020cs} (2020)  & Multi-nodes single-hop UWASN   & Spoofing {and} Eavesdropping   & Crypto-based schemes:  compressive sensing based homomorphism encryption and trust scheme &  Confidentiality and Authentication \\
  \hline
 \cite{Saeed:Access:2020} (2020) &Multi-nodes multi-hop UWASN &Wormhole & Secure and energy efficient cooperative routing & Authentication and Availability\\ 
 \hline
 \cite{khalid2020physical}  (2020) & Three-nodes (two legitimate and a malicious) LoS UWASN   & Spoofing  & Angles-of-arrival (azimuth and elevation) is used as features for physical layer authentication &  Authentication \\
  \hline
  \cite{khalid2020node} (2020)  & Two Legitimate and a malicious nodes   & Spoofing  & Maximum time-reversal resonating strength as a physical layer feature is used for authentication &  Authentication \\
  \hline
    \cite{das2020anomaly} (2020) & Multi-nodes multi-hop UWASN   & Spoofing  & Time-series data with fuzzy logic  &  Authentication\\
  \hline
   \cite{campagnaro2020replay} (2020)  & Multi-nodes multi-hop UWASN   & Spoofing (replay attack)  & Countermeasures based on time stamp and Hash value of a node &  Integrity \\
  \hline
    \cite{waqas:ett:2020} (2020) & Multi-nodes multi-hops UWASN   & Eavesdropping  & Physical layer confidentiality: resource optimization for enhancing secrecy rate of the system &  Confidentiality \\
  \hline
  \cite{Ming:WCL:2020} (2020) &Multi-nodes single-hop UWASN &Spoofing and Eavesdropping  &Secret key generation: multi-party secret key generation scheme using bi-linear mapping approach & Confidentiality and Authentication \\ \hline
  \cite{Signori:IoTJ:2020} (2020) &Two legitimate nodes with  single jammer &Jamming  & Game theoretic approach: Jammer and legitimate transmitter are considered as rational player of zero-sum game. Legitimate transmitter exploits the packet-level coding at a cost of its battery energy to protect its transmission from jamming &Confidentiality and Availability \\ \hline
  \cite{ozmen2020impact} (2020) &Multi-nodes single-hop UWASN &  Eavesdropping  & Resource allocation: two different optimization models (one minimizes eavesdropping risks while other maximizes network life time under  constraint on eavesdropping risks) are proposed to counter attacks &  Confidentiality \\
\hline 
\cite{Li:Senors:2020} (2020) & Multi-nodes multi-hop UWASN &  Eavesdropping and Active  jamming & No countermeasures are provided but machine learning algorithms are used to select optimal node for attacking &  Confidentiality and Availability  \\
\hline
 \end{tabular}
 \end{adjustbox}
 \caption{Summary of the reported security work in  underwater acoustic communication systems (2019--2020). }
  \label{tb:3B}
\end{table*}
\begin{table*}[htbp]
\centering
\begin{adjustbox}{{width=1\textwidth}}
\tiny
\begin{tabular}{ |p{1cm}|p{1.5cm}|p{1.5cm}| p{3cm}| p{2cm}|}
 \hline
\bf{Ref.} &\bf{System Architecture}  & \bf{Attacks considered} & \bf{Countermeasures proposed} &  \bf{Security Properties addressed}\\
 \hline
 \cite{signori2021geometry} (2021) &Two legitimate nodes with a jammer &Jamming &Network geometry based Game theoretical framework &Confidentiality and Availability \\ \hline
\cite{Pan:TWC:2021} (2021) & Internet of Underwater Things (IoUT) &Spoofing (Time Synchronization Attack (TSA)) & Mix-integer programming probelm is formulated and solved via expectation maximization and constrained variable threshold rounding methods& Authentication \\ \hline
\cite{Yu:ICAICA:2021} (2021) & Two legitimate and a malicious nodes  & Spoofing and Eavesdropping& Secret key based mutual authentication and symmetric key based encryption and decryption & Confidentiality, Authentication and Integrity \\ \hline
\cite{Zala:ICCCNT:2021} (2021) & Multi-nodes multi-hop UWASN &Blackhole (also known as packet-dropping attack) & Clustering with public key cryptography and challenge response mechanism  & Authentication and Availability\\ \hline
  \cite{Jain:CCPE:2021} (2021) &Multi-nodes single-hop UWASN & Spoofing (masquerade, sybil, reply, time delay) and Eavesdropping, & Symmetric-key cryptography &  Confidentiality and Authentication  \\ \hline
  \cite{Bragagnolo:COMCAS:2021} (2021) &Multi-nodes single-hop UWASN & Spoofing& Machine learning based authentication: channel taps, delay spread and received power are exploited as input features for neural network&  Authentication \\ \hline
  \cite{pelekanakis2021physical}  (2021) & Two legitimate nodes &Spoofing and Eavesdropping &Shared secret key generation: CIR based four features extraction and quantization & Confidentiality and Authentication \\ \hline
  \cite{sklivanitis2021physical} (2021) &Two legitimate and a malicious nodes &Spoofing and Eavesdropping &Shared secret key generation: Information Reconciliation and Privacy Amplification  & Confidentiality and Authentication\\ \hline
  \cite{Islam:ICEEICT:2021} (2021) &Multi-nodes multi-hop   UWASN  & Sybil &Trust management model based on Hierarchical Fuzzy System (HFS) & Authentication \\ \hline
 \cite{Zhao:ISJ:2022} (2022) &Two legitimate and a malicious nodes & Spoofing  &The estimated time-reversed CIR is convolved with
the CIRs stored in a database  &   Authentication\\
  \hline
  \cite{alharbi2022securing} (2022)&Multi-nodes multi-hop UWASN & Spoofing (specifically, depth-spoofing attacks, also known as sink-hole attacks)& Secure routing: energy-efficient depth-based probabilistic routing protocol & Authentication and Availability\\ \hline
 \end{tabular}
 \end{adjustbox}
 \caption{Summary of the reported security work in  underwater acoustic communication systems (2021--till date). }
  \label{tb:3C}
\end{table*}
\subsubsection{Security of MI based Underwater WCN}
To the best of our knowledge, there is no work reported on the vulnerabilities, attacks and countermeasures on MI based underwater WCNs. This implies that all the four properties (i.e., confidentiality, authentication, integrity and availability) are yet to be studied for MI based underwater WCN. Generally, the MI technology is claimed to be a secure wireless communication technology due to non-audible, non-visible and short range (as compared to acoustic) nature of the magnetic field. However, we believe that the predictable channel response of such networks makes it vulnerable to some attacks, e.g., eavesdropping and spoofing. Moreover, a malicious coil present in the closed vicinity of the legitimate coil may get benefit from mutual induction. 

\subsection{Security of Air-Water Wireless Communication}
\label{subsec:DMS}
\begin{figure}
\centering
\includegraphics[width=0.5\textwidth]{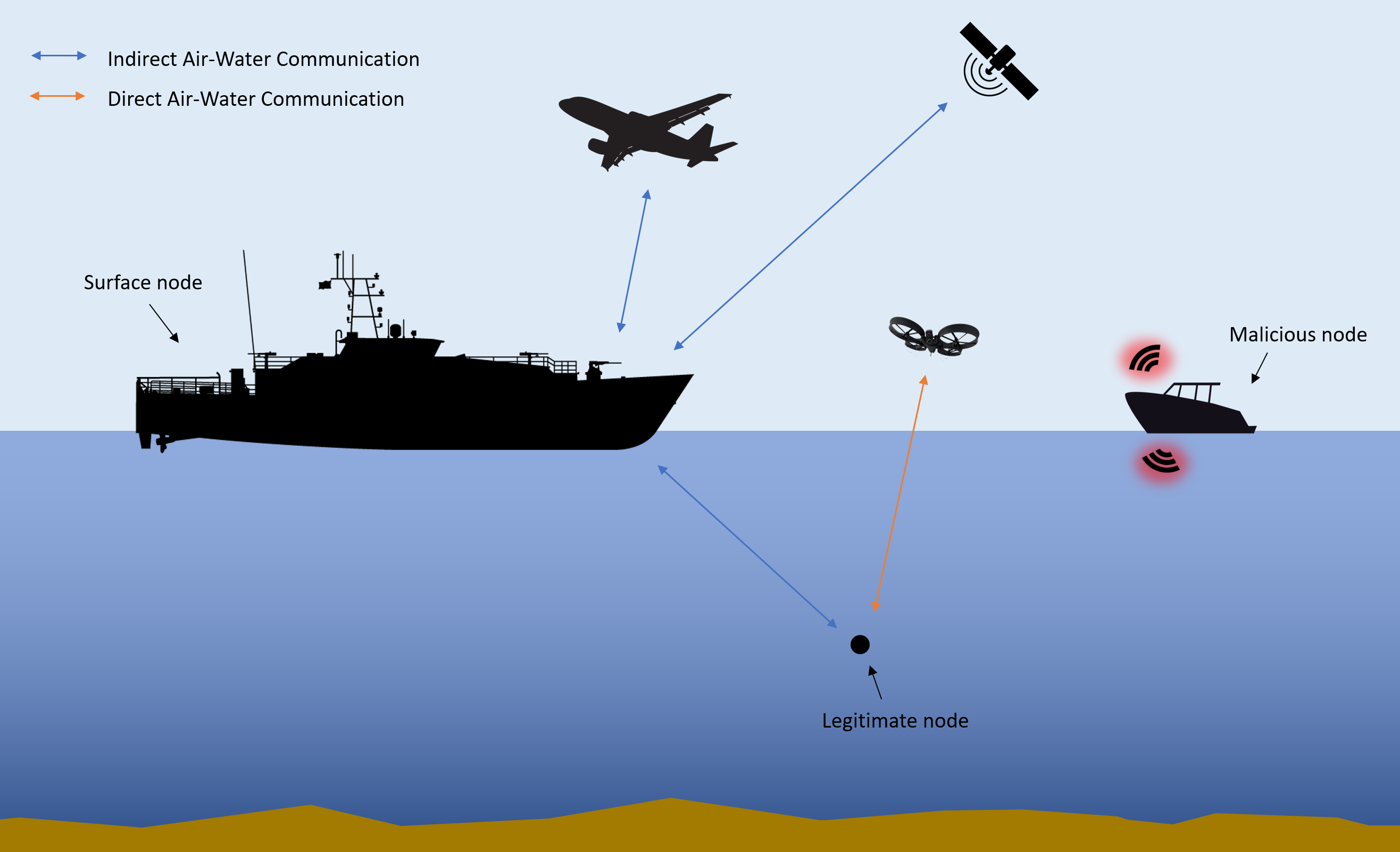}
\caption{Illustration of a scenario where a malicious node at the surface attacks the ongoing communication between underwater and airborne nodes. The malicious node can potentially lunch active and passive attacks in both mediums (however, such malicious node is not considered yet) or single medium affecting end-to-end communication. }
\centering
\label{fig:dual-medium}
\end{figure}
 
The security needs of the A-W WCNs has received very little attention by the research community so far. A typical scenario of malicious attacks in A-W WCNs is illustrated in Fig. \ref{fig:dual-medium}. There are only a handful of papers that discuss the security challenges of the indirect  \cite{Illi:Access:2018,Lou:ICL:2021,Illi:ITSC:2021,Badrudduza:Access:2021},\cite{lou:ArXiv:2020} and direct \cite{xu2018performance,guo2018channel,xie2018security,peng2022satellite} A-W WCNs, which is not surprising given that this technology is still in its infancy. The authors in \cite{Illi:Access:2018}, for the first time, study the air-water communication with RF (in the air) and optical (in water) connections. They performed secrecy outage probability analysis using end-to-end Signal to Noise Ratio (SNR) from the Low Earth Orbit (LEO) satellite to a submarine. Later, the authors in \cite{lou:ArXiv:2020} studied physical layer confidentiality with an amplify-and-forward relay node. Similarly, the authors in \cite{Lou:ICL:2021} provide physical layer confidentiality using decode-and-forward relay protocol at the surface node, while the authors in  \cite{Illi:ITSC:2021} exploit the multiple antennas at the decode-and-forward relay/surface node for physical layer confidentiality. Finally, authors in  \cite{Badrudduza:Access:2021} provide secrecy outage probability expression using generalized Gamma fading at the RF link and mixture exponential generalized Gamma fading at the underwater optical link.

Moreover, recently some theoretical study on QKD considered direct air-water wireless communication using light in both mediums\cite{xu2018performance,guo2018channel,xie2018security,peng2022satellite}. The authors in \cite{guo2018channel} proposed CV-QKD for satellite to submarine communication, where submarine is considered at a depth of less than 100m. Similarly, the authors in \cite{xie2018security}, performed Monte-Carlo simulations for CV-QKD. More recently, the authors in \cite{peng2022satellite} used measurement device independent CV-QKD to improve the results (i.e., secret key generation) of previous studies.

\section{Conclusion, Gaps and Open Problems}
\label{sec:5}

In this paper, we discussed the state-of-the-art mechanisms used for underwater and air-water wireless communication systems and further reported the existing security-related works in both mediums, mainly in terms of confidentiality, authentication, integrity and availability. The four widely used technologies (RF, optics, acoustics and MI) used in underwater and air-water wireless communication were discussed emphasizing their security related issues. We can summarize our conclusions as follows:
\begin{itemize}
    \item Despite the lack of interest in  underwater RF communication in the past due to its limitations, some recent works have given breaths to this technology.
    \item The security in underwater RF communication has not been studied yet.
    \item In underwater wireless optical communication, the literature does not contain contributions on integrity and authentication (except few theoretical works, which studied only QKD in direct light air-water wireless communication); instead, the literature seems to be mainly focusing on confidentiality.
    \item In underwater acoustic communication, confidentiality, authentication and integrity were studied with various assumptions and system models.
    \item The security or vulnerabilities of MI based underwater WCNs are yet to be studied.
    \item The literature does not contain studies on authentication and integrity in direct  air-water wireless communication systems.
    \item Few recent works on the security of direct air-water wireless communication are all theoretical and consider one configuration of air-water wireless communication (i.e., light in both mediums). 
    \item In air-water setting, no malicious node that is capable of simultaneously launching attacks in both mediums has been considered so far.
\end{itemize}

Despite the existing literature on the security of underwater and air-water wireless communication systems and previously mentioned research gaps, there are still many open problems, which we briefly discuss in the following subsections.


\subsection{Smart Malicious nodes}
A smart malicious node is a node that can assess the situation for launching attacks, for example a smart malicious node can be a node that can switch between passive and active modes on the fly \cite{Wang:WCSP:2017,Xu:TVT:2018}. Therefore, an interesting problem is to analyze the existing state-of-the-art underwater communication technologies under smart malicious nodes. In particular, it will be interesting to conduct such security analysis in the presence of malicious nodes equipped with the state-of-the-art machinery (i.e., Multiple Input and Multiple Output (MIMO), full duplex, GPUs capable of launching attacks in both mediums). The authors in a recent work \cite{Long:AIP:2021} developed
 a full-duplex system using thermoacoustic effect and microwave vibration measurement for direct air-water wireless communication system using 10cm dipped water node and 30cm airborne node. This can naturally be extended to investigating security aspects of such communication. Additionally,
 %
  to the best of our knowledge, a full-duplex malicious node is not considered in the literature.  
On the other hand, active attacks, such as jamming and spoofing, and passive attacks, such as eavesdropping, on direct air-water wireless communication are not explored yet. Hence, we believe that studying the security aspect of direct air-water wireless communication should be considered for future work. Similarly, we also need to account for malicious nodes capable of listening in both mediums in air-water wireless communication. In fact, this area lacks formal treatment of threat modelling to capture the malicious node's wide range of behaviors, strategies, the computing and communication facilities. The literature so far misses contributions on modeling of the malicious nodes themselves. 

\subsection{Impact of Security mechanisms on other system parameters}

It is worth noting that adding security of a system will introduce an unavoidable overhead and implications. Since the overall performance of any system is a tangled mix of various system parameters, it is important to study the impact of security on relevant performance metrics, such as energy efficiency, throughput and reliability.

Indeed, some works discussed such trade-off between security and performance on wireless communications systems \cite{Yan:ICC:2014, Yu:ISJ:2018, zhong:JSAC:2018, Yu:TWC:2019,Henrik:Globcom:2017,Henrik:JSAC:2019, aman2020effective}. In \cite{aman2020effective}, the authors provided relation of effective capacity and authentication parameters (false alarm and missed detection) as a closed-form expression in underwater acoustic communication, where they investigated the impact of physical layer authentication on effective capacity (a link layer metric that assesses the reliability level offered by a system).

%
Similarly, we believe it is necessary to thoroughly investigate the trade-off between security mechanisms and throughput, energy, spectral efficiencies of both underwater and air-water wireless communication systems in order to assess the cost of enforcing a security mechanism, given a desired level of security.

\subsection{AI/Machine Learning-empowered Security solutions}
Recently, machine learning tools have been leveraged to improve the performance of wireless communication networks \cite{sun:CST:2019,gunduz:JSAC:2019,morocho:Access:2019,Zhu:ICM:2020}. From security perspective in the air-water communication systems, a recently reported work \cite{waqas:ett:2020} used trained artificial neural networks to combat eavesdropping in underwater acoustic communication. Similarly, in underwater acoustic sensor networks, there has been an interest in using machine learning tools for, e.g., authentication \cite{Xiao:ICL:2018, Bragagnolo:COMCAS:2021} and counter jamming \cite{xiao:ICLJamming:2018}. As security improvements provided by machine learning tools in underwater communication looks quite promising, we believe that experimenting on air-water communication will most likely yields equally interesting results. 

\subsection{Cross-layer Security}
To further enhance the security and privacy of a communication network, cross-layer approaches have been used where security is provided at different protocol stack layers \cite{sharma:2011:cross,Farag:sgrid:2014,Wurm:TMSCS:2017,sumalatha:JAIHC:2021}. Unlike the traditional security approaches where security is provided at a particular layer, cross layer approaches study security in multiple layers and protects information as it crosses the layers.  We believe that such approaches have not been studied in air-water communication systems and can potentially improve the security in these systems. 





\subsection{Secure Localization}
Localization continues to be an important topic for underwater nodes because of the unavailability of  Global Positioning System (GPS) signals underwater. Instead, some reference/anchored nodes were used to localize deployed unknown underwater nodes  \cite{tan:OE:2011,Inam:Access:2019,Inam2:Access:2019,saeed:AHN:2019,su:JS:2020}. However, an open problem is to investigate localization of underwater nodes in direct air-water systems, possibly by complementing both reference/surface underwater nodes, and traditional localization mechanisms from over water. 
On the other hand, recently reported localization attacks, such as Sybil and Wormhole \cite{goyal2015wormhole,sharma2017classification,kaliyar2020lidl} imply that there is a great need to develop secure localization schemes. Some efforts have been made in this direction  \cite{Li:ICM:2015,baranidharan:ICTACT:2018,shanthi:ICAECC:2018, YANG:phycom:2019,misra:TAAS:2021}, but more investigations are needed, particularly, for emerging communication technologies in air-water communication. 


\section*{Acknowledgement}
This work is partially funded by the G5828 ``SeaSec: DroNets for Maritime Border and Port Security" project under the NATO's Science for Peace Programme.

\footnotesize{
\bibliographystyle{IEEEtran}
\bibliography{references}
}

\vfill\break

\end{document}